\documentclass[preprint,5p,times, authoryear, sort, comma, singleblind, nopreprintline]{elsarticle}
\usepackage[utf8]{inputenc}
\usepackage{amsmath,amsfonts,amssymb, mathtools, blindtext}
\usepackage{natbib}
\usepackage{multirow}
\usepackage{subfigure}
\usepackage{graphicx}
\usepackage{tabularx}
\usepackage{amssymb}
\usepackage{float}
\usepackage{csvsimple} 
\usepackage{xcolor}
\usepackage{appendix}
\usepackage{subcaption}   
\usepackage[font=small,labelfont=bf]{caption}  
\usepackage{array}        
\usepackage{booktabs}     
\usepackage{hyperref}
\hypersetup{
    colorlinks=false,
}

\graphicspath{ {./experiment-figs/} }

\begin{document}

\begin{frontmatter}

\title{An Analytics-Driven Approach to Enhancing Supply Chain Visibility with Graph Neural Networks and Federated Learning}

\author[inst1]{Ge Zheng}
\author[inst1,inst2]{Alexandra Brintrup\corref{cor1}}
\affiliation[inst1]{organization={Institute for Manufacturing, University of Cambridge},country={United Kingdom}}
\affiliation[inst2]{organization={Alan Turing Institute, London},country={United Kingdom}}
\cortext[cor1]{Corresponding author: ab702@cam.ac.uk}

\begin{abstract}

In today’s globalised trade, supply chains form complex networks spanning multiple organisations and even countries, making them highly vulnerable to disruptions. 
These vulnerabilities, highlighted by recent global crises, underscore the urgent need for improved visibility and resilience of the supply chain. 
However, data-sharing limitations often hinder the achievement of comprehensive visibility between organisations or countries due to privacy, security, and regulatory concerns. 
Moreover, most existing research studies focused on individual firm- or product-level networks, overlooking the multifaceted interactions among diverse entities that characterise real-world supply chains, thus limiting a holistic understanding of supply chain dynamics.
To address these challenges, we propose a novel approach that integrates Federated Learning (FL) and Graph Convolutional Neural Networks (GCNs) to enhance supply chain visibility through relationship prediction in supply chain knowledge graphs. 
FL enables collaborative model training across countries by facilitating information sharing without requiring raw data exchange, ensuring compliance with privacy regulations and maintaining data security. 
GCNs empower the framework to capture intricate relational patterns within knowledge graphs, enabling accurate link prediction to uncover hidden connections and provide comprehensive insights into supply chain networks.
Experimental results validate the effectiveness of the proposed approach, demonstrating its ability to accurately predict relationships within country-level supply chain knowledge graphs. 
This enhanced visibility supports actionable insights, facilitates proactive risk management, and contributes to the development of resilient and adaptive supply chain strategies, ensuring that supply chains are better equipped to navigate the complexities of the global economy.
\\
\newline SDG 9: Industry, innovation and infrastructure

\end{abstract}


\end{frontmatter}


\section{Introduction} \label{sec: introduction}

In the context of today's highly interconnected global economy, supply chains extend across multiple regions and countries, forming complex networks of suppliers, manufacturers, distributors, and customers \citep{mangan2016global}. 
These intricate networks have proven to be highly susceptible to disruptions, as evidenced by recent global crises such as the COVID-19 pandemic and the Ukraine war, which severely disrupted production and distribution systems worldwide \citep{ivanov2020viability}. 
Additionally, the semiconductor shortage has further illustrated the vulnerabilities of global supply chains, significantly affecting industries ranging from automotive manufacturing to consumer electronics \citep{ramani2022understanding}. 
Famous companies, such as Apple and Toyota, have experienced substantial challenges due to these disruptions, underscoring the critical importance of enhancing supply chain visibility and building resilience to mitigate the impacts of such events \citep{pettit2010ensuring, ivanov2020predicting}

Enhanced supply chain visibility allows operators and managers to mitigate risks, optimise operational efficiency, and ensure compliance with regulatory frameworks \citep{wang2016big}. 
For example, manufacturers can proactively address potential disruptions in the supply of raw materials, while retailers can effectively manage inventory levels over fluctuating demand and supply uncertainties \citep{larson2001designing}. 
Despite its advantages, achieving supply chain visibility presents significant challenges. 
One major obstacle is the limited availability of data, as companies often hesitate to share proprietary information due to competitive pressures and strict data privacy regulations such as the General Data Protection Regulation (GDPR) in the European Union \citep{gdpr2016general}. 
Furthermore, data heterogeneity complicates integration efforts, with supply chain data varying widely in format, quality, and granularity between different organisations and countries \citep{chae2015insights}. 
Besides, cross-border information sharing is also constrained by policy barriers and security concerns, creating additional hurdles for global supply chain coordination \citep{colicchia2019information}.

Traditional approaches to modelling supply chains for visibility typically pay attention on firm- or product-level networks, focusing on the isolated relationships between companies or products \citep{harland2001taxonomy, kim2012product}. 
While these models offer valuable insights into specific components of the supply chain, they often fail to consider the multifaceted interactions among diverse entities, including customers, companies, products, and certifications \citep{hearnshaw2013complex}. 
This narrow focus limits the ability to understand and predict the intricate dynamics of supply chains, where the interplay between various entities is crucial to achieving comprehensive visibility \citep{paulheim2017knowledge}.

Representing supply chains as knowledge graphs offers a holistic and adaptable approach to modelling the entire ecosystem \citep{kosasih2024towards}. 
Unlike traditional methods, knowledge graphs integrate diverse entities and relationships into a unified framework, effectively capturing the rich semantics and intricate interdependencies natural in real-world supply chains. 
By enabling link prediction, knowledge graphs facilitate the discovery of hidden relationships and the anticipation of emerging connections among various entities. 
This advanced approach addresses the limitations of firm- or product-level networks by encompassing the broader context and accounting for the complex interactions between multiple types of entities and relationships, thus providing deeper insights into supply chain dynamics.

For example, predicting a new ``has\_cert'' relationship between a company and a certificate can help identify compliance risks or quality issues that might not be apparent when analysing firm-level data alone. 
Similarly, uncovering a potential ``buys'' relationship between a customer and a product can reveal emerging market trends and customer preferences that are critical for strategic decision-making. 
By leveraging the comprehensive and interconnected structure of supply chain knowledge graphs, organisations can gain deeper insights and make more informed decisions.

However, achieving this level of supply chain visibility remains significant challenges such as limited data availability, heterogeneity in data formats, policy barriers, and privacy concerns. 
To address these challenges, we propose a new approach that leverages federated learning (FL) and graph convolutional neural networks (GCNs) to predict relationships within supply chain knowledge graphs. 
FL facilitates collaborative training of a shared machine learning model across multiple parties without requiring the exchange of raw data, thereby ensuring data privacy and compliance with regulatory constraints. 
Each country or organisation trains the model locally on its proprietary supply chain data, sharing only model parameters or updates for aggregation. 
The model architecture leverages GCNs to capture the intricate relational structures within supply chain networks, effectively modelling the diverse entities and relationships embedded in the knowledge graph \citep{kipf2016semi}.

Our proposed methodology addresses several key limitations of existing approaches. 
First, it prioritises data privacy and security by eliminating the need for raw data exchange, ensuring that sensitive information remains protected. 
Second, it effectively manages data heterogeneity by allowing each participant to retain their local data formats and schemas, while the federated model learns a unified representation across diverse datasets. 
Third, it ensures compliance with policy and regulatory requirements by keeping data and processing within local servers. 
Finally, by incorporating advanced machine learning techniques such as graph convolutional neural networks (GCNs), the methodology significantly enhances the accuracy and robustness of link prediction within the knowledge graph, ultimately improving supply chain visibility and supporting more informed decision-making.

The rest of the paper is as follows. 
Section~\ref{sec: literature} provides a comprehensive review of existing studies on supply chain visibility, link prediction, and federated learning. 
Section~\ref{sec: methodology} details the proposed approach for link prediction within the constructed supply chain knowledge graph. 
Section~\ref{sec: case study} presents a real-world case study consisting of the data employed, experimental settings, benchmarks for comparison, and a discussion of the results. 
Finally, Section~\ref{sec: conclusion} summarises the findings, discusses managerial implications, highlights limitations, and offers directions for future research.

\section{Related Works} \label{sec: literature}

\subsection{Supply Chain Visibility}

Supply chain visibility refers to the ability of stakeholders to access accurate, complete, and timely information regarding the movement of materials and products throughout the supply chain \citep{saint2016supply}. 
This includes the flow of raw materials, components, and finished goods from suppliers to manufacturers, through distributors and retailers, and ultimately to end customers.
The growing complexity of supply chains, driven by globalisation and the increasing need for agility and resilience, has highlighted the importance of visibility \citep{baycik2024quantitative}. 
Enhanced visibility allows organisations to optimise processes, reduce operational costs, improve inventory management, and mitigate risks associated with disruptions by providing real-time insights into the movement of goods, materials, and information. 
During the COVID-19 pandemic, for example, companies such as Apple \citep{singhkang2023apple}, Walmart, and Amazon \citep{jain2021overview} showcased superior adaptability to rapidly changing conditions due to their robust supply chain visibility, which allows them to monitor and adjust operations in real-time. 
They maintain supply chain continuity even under challenging circumstances by leveraging advanced technologies for tracking, analytics, and communication. 
This underscores the critical role of supply chain visibility in enabling swift responses to disruptions and maintaining efficient operations.

Achieving supply chain visibility presents significant challenges, requiring multifaceted approaches that leverage advanced technologies and standardised systems. 
Key enablers include technologies such as the Internet of Things (IoT), Radio-Frequency Identification (RFID), Global Positioning Systems (GPS), and advanced data analytics tools, which facilitate real-time data collection and analysis across the supply chain. 
Furthermore, achieving seamless information sharing among stakeholders necessitates data integration and standardisation through unified platforms and protocols, such as Electronic Data Interchange (EDI) and Application Programming Interfaces (APIs).

Current methods for enhancing supply chain visibility include the deployment of Enterprise Resource Planning (ERP) systems \citep{jeyaraj2010implementation}, RFID \citep{ilic2009increasing}, IoT \citep{pundir2019improving}, and blockchain technologies \citep{mukhtar2020blockchain}. 
These solutions enable end-to-end visibility by collecting and sharing data across the supply chain network. 
For example, RFID and IoT devices facilitate real-time tracking of goods \citep{jeyaraj2010implementation, ilic2009increasing}, while blockchain technology ensures data integrity and transparency through its decentralised and immutable architecture \citep{mukhtar2020blockchain, foley2023bayes}.

Although these approaches demonstrate considerable potential, they still face certain limitations. 
ERP systems often struggle with integrating data from diverse sources and may fail to deliver real-time visibility due to delays in data processing. 
Similarly, while RFID technology enables tracking and data collection, its adoption can be expensive and unaffordable for small and medium-sized enterprises (SMEs), and its utility is reduced in the absence of standardised data-sharing protocols. 
Blockchain technology, although promising for decentralised data sharing, faces challenges related to scalability, high energy consumption, and the need for universal agreement among participants to adopt a common platform. 
These limitations highlight the need for continuous innovation and the development of more inclusive and scalable solutions to enhance supply chain visibility.

The emergence and widespread adoption of Artificial Intelligence (AI) including Machine Learning (ML), Deep Learning (DL) and Generative AI (GenAI) technologies across various sectors have introduced promising solutions to address the limitations of traditional supply chain visibility approaches. 
These technologies enable advanced capabilities such as predictive analytics, anomaly detection, and decision support systems that process and analyse large volumes of heterogeneous data \citep{silva2017improving}.
AI can enhance visibility by automating data collection, processing unstructured information, and generating insights that are difficult to achieve with conventional methods. 
Moreover, it can also contribute to improved supply chain management by predicting demand fluctuations \citep{kochak2015demand}, optimising route planning for logistics vehicles \citep{hu2020optimal} and supplier selection for efficiency \citep{modares2025bayesian}, and identifying potential disruptions before they occur \citep{brintrup2020supply}. 
These capabilities collectively strengthen supply chain visibility and operational resilience.

While AI offers substantial benefits for enhancing supply chain visibility, its implementation still faces significant challenges.
A primary concern involves data privacy and security, particularly when information must be shared across organisational and national boundaries.
Regulatory frameworks, such as GDPR \citep{gdpr2016general}, impose stringent requirements on data sharing, necessitating the development of innovative methodologies that leverage AI capabilities while ensuring compliance and protecting sensitive information.

\subsection{Link Prediction for Supply Chain Visibility}

Link prediction is a common but important problem in network analysis, focusing on the identification of missing or hidden links and the prediction of future connections between entities based on observed patterns \citep{martinez2016survey}. 
In the context of supply chains, entities such as suppliers, manufacturers, products, and customers, along with their relationships, can be represented and structured as knowledge graphs \citep{kosasih2024towards}. 
Applying link prediction to these graphs can enable the identification of missing and potential relationships among entities, thus enhancing supply chain visibility \citep{kosasih2022machine, brockmann2022supply} and supporting informed decision-making.

The importance of this approach has been highlighted by disruptions caused by the COVID-19 pandemic, which exposed vulnerabilities in the supply of critical items such as personal protective equipment (PPE) and medical devices. 
Hospitals and governments experienced shortages partly due to limited awareness of the complex supplier networks involved in the production of these items, such as the dependency of N95 masks on specialised materials like melt-blown fabric, which is produced by a small number of suppliers. 
By predicting potential partnerships between suppliers, materials, and manufacturers, link prediction can facilitate the identification of alternative sourcing opportunities, improving supply chain resilience and adaptability.
Link prediction in supply chain knowledge graphs offers a comprehensive and detailed perspective on the supply chain network. 
By leveraging this capability, organisations can gain deeper insights into their market positions, enhance risk management strategies, optimise operational efficiency, and make more informed decisions to strengthen overall supply chain performance.

Approaches to link prediction in knowledge graphs primarily include methods such as matrix factorisation \citep{nickel2011three}, random walk-based algorithms \citep{lao2010relational}, and, more recently, deep learning techniques like Graph Convolutional Neural Networks (GCNs) \citep{kipf2016semi}. 
Matrix factorisation methods, such as those developed by \citep{nickel2011three}, decompose the adjacency tensor of a knowledge graph to uncover latent interactions between entities and relationships. 
For instance, matrix factorisation can identify associations such as the capital-city relationship in a graph containing entities like ``London'' and ``United Kingdom''. While effective at capturing linear patterns, this approach struggles with modelling complex, non-linear relationships inherent in knowledge graphs with multiple entities and relationships.
Random walk-based algorithms \citep{lao2010relational} utilise random walks across the graph to discover meaningful paths between entities, enabling the prediction of unobserved links. 
These algorithms can capture longer-range dependencies compared to matrix factorisation; however, they are computationally intensive and lack the ability to generalise effectively to unseen data.

To address the limitations of earlier methods, deep learning techniques such as Convolutional Neural Networks (CNNs) and Graph Convolutional Neural Networks (GCNs) have been introduced for link prediction tasks. 
For example, \citep{dettmers2018convolutional} applied 2D convolutions to entity embeddings to model complex interactions in social networks, while \citep{kipf2016semi} developed GCNs to aggregate feature information from neighbours of a node, effectively capturing both local and global graph structures. 
Unlike CNNs, which focus on local interactions and cannot leverage the graph's structural information, GCNs excel in exploiting the topology of knowledge graphs. 
In a citation network, for instance, a GCN can learn representations of research papers by considering not only direct citations but also the broader citation context, enabling more accurate predictions of future citation links.

Although link prediction has been widely studied across domains such as social networks, biological networks, and recommendation systems due to its ability to address various problems, its applications in the supply chain sector have received limited attention. 
Few existing research studies on link prediction in supply chains can be broadly classified into two categories: approaches that implement link prediction on supply chain graphs with single-type relationships and those that extend the problem to supply chain knowledge graphs with multiple types of entities and relationships.

For supply chain graphs with single-type relationships, \citep{brintrup2018predicting} proposed a machine learning-based link prediction approach to identify supplier interdependencies within a manufacturing supply network. 
Similarly, \citep{xie2019higher} utilised the alternative direction multiplier method (ADMM) to predict potential links between nodes in an enterprise network, aiming to recommend high-quality partnerships. 
In another study, \citep{kosasih2022machine} employed Graph Neural Networks (GNNs) to infer previously unknown links in an automotive supply chain network.
In contrast, other studies have framed link prediction as a problem of reconstructing production networks. 
For example, \citep{ialongo2022reconstructing} developed a generalised maximum entropy reconstruction method to infer firm-level networks using partial data, showcasing the applicability of link prediction in network reconstruction. 
Similarly, \citep{mungo2023reconstructing} applied a gradient-boosting approach to uncover hidden relationships and rebuild production networks effectively.

In the context of supply chain knowledge graphs, which extend link prediction to multiple types of relationships, recent advancements have built upon earlier work in single-type graphs. For example, \citep{brockmann2022supply} employed GNNs to predict links in an uncertain supply chain knowledge graph. 
Building further on these efforts, \citep{kosasih2024towards} introduced a neurosymbolic machine learning approach that integrates GNNs with knowledge graph reasoning to predict multiple types of links across two supply chain networks, demonstrating a significant step forward in enhancing supply chain visibility and understanding.

A shared limitation across these studies is their reliance on information sharing between firms. 
Due to concerns over data privacy, security, and competitive advantages, companies are often reluctant to disclose information about their supply chain relationships. 
Moreover, global supply chains frequently involve cross-border information sharing, which is further constrained by country-specific policies and regulations. 
These challenges highlight a critical question: \textit{How can supply chain visibility be achieved without requiring the exchange of raw data, leveraging link prediction methods as an alternative?}

Despite its importance, this question has not been addressed in the existing literature, leaving a significant research gap. 
This study aims to bridge this gap by developing a novel link prediction approach that enhances supply chain visibility while eliminating the need for direct raw data sharing between firms. 
Our approach is designed to maintain data privacy and security constraints, providing a pathway to achieve supply chain transparency without compromising sensitive information.

\subsection{Federated Learning for Link Prediction}

Federated learning (FL), originally developed by the Google AI team \citep{mcmahan2017communication}, is a machine learning framework designed to address data privacy, security, and communication efficiency issues in decentralised systems. 
It enables the training of a shared global machine learning model across multiple devices or servers without requiring the transfer of local data.
In this framework, a shared global machine learning model is initialised on a central server and distributed to local clients. 
Each client trains the model on its local dataset and returns the updated model parameters to the central server where the updated parameters are aggregated to refine the shared global model. 
This iterative process continues until the shared global model converges, resulting in a final shared global model learned from all local datasets while maintaining data privacy. 
FL offers a solution to issues associated with centralised data processing, including risks of data breaches and compliance with regulations that restrict data sharing. 
By holding data on local devices, it ensures privacy preservation and reduces communication overhead, facilitating collaborative learning without compromising data security.

Given its obvious advantages, FL has been successfully applied across diverse domains. 
In digital health \citep{rieke2020future}, FL enables multiple healthcare institutions to collaboratively train models while safeguarding sensitive patient data. 
In wireless communication \citep{niknam2020federated}, it enhances network optimisation by preserving user privacy and reducing the need to transmit raw data to central processors. 
Similarly, in the Internet of Things (IoT) \citep{khan2021federated}, FL allows edge devices to improve learning models without centralising data. 
In vehicular networks \citep{elbir2022federated}, FL facilitates connected vehicles in sharing insights to enhance 3D object detection for autonomous driving, minimising data transmission requirements.

Meanwhile, FL has also been explored in various domains for link prediction, demonstrating its capability to address privacy concerns in decentralised environments. 
For example, \citep{jiang2020federated} developed an FL-based link prediction model that enables multiple social platforms to collaboratively predict potential user connections without sharing sensitive data. 
Similarly, \citep{jiang2022fedncf} applied FL to train machine learning models on highly decentralised data for recommendation systems. 
In the biomedical field, \citep{rieke2020future} utilised FL to predict associations between diseases and genes in distributed biomedical networks across hospitals and research institutes. 
These applications highlight the effectiveness of FL in facilitating link prediction while maintaining data privacy and security.

Given its success in various domains for link prediction, FL shows significant potential for application in supply chain networks, particularly in global supply chains characterised by decentralisation and diverse entities, such as suppliers, manufacturers, and distributors, operating under strict confidentiality due to competitive and regulatory concerns. 
FL can enhance supply chain visibility by predicting missing or hidden relationships within the network without compromising data privacy. 
By applying FL to link prediction, organisations can collaboratively uncover hidden dependencies and construct comprehensive supply chain knowledge graphs, thus improving risk management, demand forecasting, and strategic decision-making. 
This approach not only aligns with the decentralised nature of supply chain networks but also provides a robust solution for achieving global supply chain visibility through collaborative learning without requiring raw data sharing.

Despite its promising prospects, the application of FL for link prediction in supply chains, particularly to improve visibility, remains underexplored. 
This study aims to address this gap by investigating the use of FL for link prediction on supply chain knowledge graphs, enabling enhanced visibility while preserving data privacy and competitive constraints.

\section{Graph Convolution Neural Network-based Federated Learning Framework} \label{sec: methodology}

\subsection{Preliminaries} \label{subsec: definition of SCKG}

\begin{figure}[th!]
    \centering
    \includegraphics[width=0.35\textwidth]{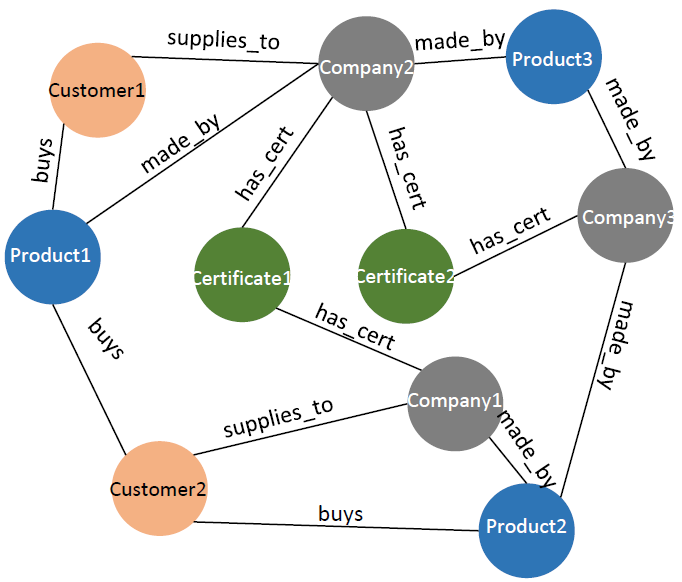}
    \caption{A supply chain knowledge graph example constructed by a part of UK data.}
    \label{fig: knowledge graph example.}
\end{figure}

A supply chain knowledge graph (SCKG) is a specialised data structure that represents entities and their relationships within a supply chain network, offering a comprehensive view of its complexities. 
As depicted in Figure~\ref{fig: knowledge graph example.}, an example SCKG maps connections such as manufacturing, supplying, purchasing, and certification relationships between entities. 
The graph illustrates that \texttt{Company2} manufactures both \texttt{Product1} and \texttt{Product3}, supplies \texttt{Product1} to \texttt{Customer1}, and shares \texttt{Certificate1} with \texttt{Company1}, indicating compliance with the same standards or regulations.

By leveraging an SCKG, companies can gain valuable insights into the end-to-end flow of goods, from production to final delivery, and proactively identify vulnerabilities within the supply chain. 
For example, if a supplier providing a critical product also serves competitors, any disruption affecting that supplier could have cascading impacts across multiple parts of the supply chain, even the entire sector. 
SCKGs thus play a vital role in enhancing visibility, supporting risk management, and enabling informed decision-making within complex supply chain networks.

Given a common graph, \(G=(V, E)\) \citep{shen2022comprehensive}, in which $V$ represents a collection of nodes and $E$ donates a set of edges, it can be extended to define an SCKG as:

\[G=(V,E,T,R,\phi,\psi)\] 

\noindent where $V=\{v_1, v_2, v_3, \dots, v_n\}$ is the set of nodes categorised into types $T=\{t_1, t_2, t_3, \dots, t_t\}$, such as companies, customers, products, and certificates. 
$E \subseteq (V \times V)$ is the set of edges, each $e=(v_i, v_j)$ representing a relationship between nodes $v_i$ and $v_j$. 
$R=\{r_1, r_2, r_3, \dots, r_m\}$ denotes the set of relationship types that describe the nature of interactions between nodes, such as ``\textit{supplies\_to}'', ``\textit{made\_by}'', ``\textit{buys}'' and ``\textit{has\_cert}''. 
The function $\phi$ assigns a relationship type $r \subseteq R$ to each edge $e \subseteq E$ while $\psi$ assigns a type $t \subseteq T$ to each node $v \subseteq V$. 
With this definition, an SCKG can be constructed as a set of triples, where each triple consists of two nodes $v_i$ and $v_j$ connected by a relationship $r$, expressed as $(v_i, r, v_j)$. 
Notably, this formalism is not restricted to SCKGs but is also applicable to other types of knowledge graphs, making it an unified framework for representing complex networks.

Thus, the link prediction problem on the supply chain knowledge graph can be formalised as estimating the likelihood $P(e')$ that an edge $e=(v_i, v_j)$, representing a relationship of type $r$, should exist between two nodes $v_i$ and $v_j$, based on the existing structure of the knowledge graph $G$.

\subsection{GCN-based Federated Learning Framework}

\begin{figure*}[th!]
    \centering
    \includegraphics[width=0.85\textwidth]{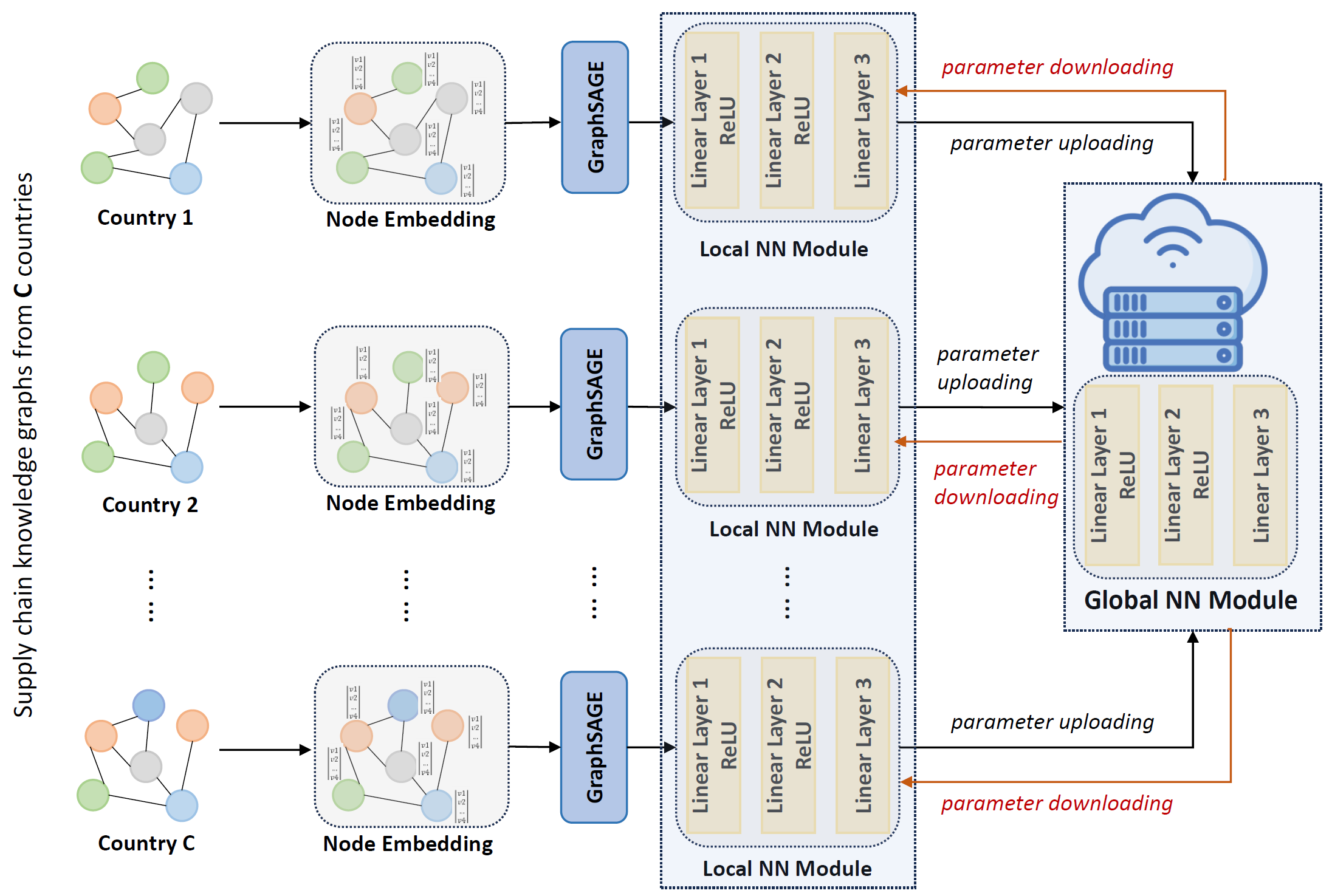}
    \caption{The developed framework of graph convolution neural network-based federated learning}
    \label{fig: GCN-FL framework}
\end{figure*}

Figure~\ref{fig: GCN-FL framework} shows our federated learning framework, which leverages GraphSAGE \citep{hamilton2017inductive} for decentralised learning across multiple supply chain knowledge graphs (SCKGs). 
This approach enables a shared model training for link prediction on different SCKGs, without requiring the exchange or centralisation of sensitive or private raw data. 
The functionality of GraphSAGE will be further detailed in Section~\ref{subsec: GraphSAGE}.

The proposed framework involves $C$ SCKGs, each representing the supply chain knowledge of a different country. 
These SCKGs comprise entities such as companies, products, and certificates, as well as relationships such as ``\textit{buys}'', ``\textit{supplies\_to}'', ``\textit{made\_by}'', and ``\textit{has\_cert}'', which describe the flow of goods and services. 
Given privacy and security constraints that prevent direct data sharing, the FL framework facilitates collaboration among these countries by enabling the training of a global model without requiring raw data sharing. 
Each country retains its SCKG locally.

The entire learning process comprises \textit{four steps}, detailed as follows:

\textit{Step 1} involves data preparation for the GCN-based FL training. 
Each country’s SCKG is prepared for FL through data collection, processing, and integration, as described in Section~\ref{sub: data description}.
The processed data are then structured into an SCKG, following the example SCKG provided in Section~\ref{subsec: definition of SCKG}.

\textit{Step 2} focuses on node embedding and representation learning of SCKGs using GraphSAGE. 
The constructed SCKG is input into a node embedding module, which maps the identities of entities into dense vector embeddings. 
Each entity is represented as a learnable vector of fixed dimensions. 
These embeddings are then processed by GraphSAGE, which aggregates information from neighbouring nodes to update the embeddings iteratively. 
This aggregation mechanism enables the model to capture both local and structural context, making it suitable for decentralised graph learning.

\textit{Step 3} refers to the local neural network (NN) module training for parameter sharing in \textit{Step 4}. 
Given the heterogeneity in graph structures, node features, and relationship types across different countries’ SCKGs, the GraphSAGE layers trained on one country’s data may not generalise well to another’s. 
For example, a company in one country may have entirely different suppliers or products compared to another. 
To address this, an NN module comprising three fully connected layers, each followed by the non-linear activation function ``ReLU'', is appended after the GraphSAGE layers. 
This NN module is more generalisable and serves as the shared component for global model aggregation.
By keeping the GraphSAGE layers local to each country, the model can adapt specifically to the unique structure of each SCKG, while the parameters of the NN module are shared for global learning. 
This separation ensures that the federated model remains robust to local graph variations. 
Each country learns high-quality node embeddings tailored to its SCKG, while the global model benefits from aggregated knowledge encoded in the shared layer parameters.

\textit{Step 4} involves global model aggregation.
In this step, the parameters of the local NN modules from all $C$ countries are aggregated to generate a global NN module. 
The aggregation is performed using an averaging method, ensuring that contributions from all countries are integrated equally. 
The updated global model is then distributed back to each country, where it is fine-tuned locally using the respective SCKG. 
This iterative process continues over multiple cycles, progressively enhancing the accuracy and generalisability of the global model.

\subsection{GraphSAGE} \label{subsec: GraphSAGE}

In the FL framework, we employ GraphSAGE \citep{hamilton2017inductive} to learn node representations within SCKGs. 
GraphSAGE is particularly suitable for our use case as it supports inductive learning, enabling the generation of embeddings for nodes not present in the training data. 
In contrast, traditional GCNs operate in a transductive setting, requiring retraining whenever new nodes are introduced. 
This is especially relevant in SCKGs, where entities such as new companies or products frequently emerge, and existing entities may disappear due to changes in business dynamics. 
GraphSAGE allows for the prediction of new relationships (e.g., when a new company joins the SCKG by purchasing an existing product) without necessitating model retraining. 
Besides, GraphSAGE employs a neighbourhood aggregation method that learns node embeddings by sampling and integrating information from a node’s local neighbourhood.

In this framework, GraphSAGE acts as the node representation learner following the node embedding module that initialises each node's features as a vector. 
For example, the initial features of a node $v_i$ are represented as $h^0_{v_i}$. 
GraphSAGE then operates in three stages: \textit{neighbourhood sampling}, \textit{aggregation of neighbour information} and \textit{node representation updating}.

\textit{Neighbourhood sampling} involves selecting a fixed number of neighbours for each node, rather than using the entire neighbourhood, which may be computationally expensive for large graphs. 
For a given node $v_i$, its neighbours are denoted as $N(v_i)$, and GraphSAGE samples $k$ neighbours from $N(v_i)$. 
For example, \texttt{company A} supplies products to a certain number of customers, and only the information from $k$ sampled customers is used to update the representation of \texttt{company A}. 
This sampling approach ensures the scalability of the model to large knowledge graphs by avoiding exhaustive neighbour processing.

\textit{Aggregation of neighbour information} is the next step aiming to aggregate feature information from the selected $k$ neighbours of each node. 
In the original GraphSAGE implementation \citep{hamilton2017inductive}, three aggregation methods were introduced: mean aggregator, pooling aggregator, and LSTM aggregator. 
In our framework, we use the mean aggregator for its simplicity and computational efficiency, as it computes the average of the neighbours’ embeddings. 
This method is well-suited for SCKGs, which consist of diverse relationship types (e.g., \textit{supplies\_to}, \textit{buys}, \textit{made\_by}, and \textit{has\_cert}). 
The mean aggregator avoids overfitting to specific relationship patterns and maintains a balanced summary of neighbourhood information.
The aggregation is computed as:

\[h^{(k)}_{v_i} = ReLU\Bigg(W \cdot mean \Big({h^{(k-1)}_{v_i}} \cup \{{h^{(k-1)}_u}: {u \in N(v_i)}\}\Big)\Bigg)\]

\noindent where $W$ is a learnable weight matrix, $mean$ denotes the averaging operation over the node $v_i$ and its neighbours $u$, and $ReLU$ introduces non-linearity into the model.

\textit{Node representation updating} for a node $v_i$ after $k$ aggregation steps is denoted as $h^{(k)}_{v_i}$, which combines the features of $v_i$ and its neighbours up to $k$ hops away.
This process is repeated for $k$ layers, enabling each node to incorporate information from an increasingly larger local neighbourhood. 
Once node embeddings have been updated through the aggregation process, they are passed to an NN model for further feature learning and relationship prediction.

Given the final representations for two nodes $v_i$ and $v_j$, $h^{(k)}_{v_i}$ and $h^{(k)}_{v_j}$, the likelihood of a relationship between these nodes is computed using a dot product followed by a sigmoid activation function:

\[\hat{y}_{v_i; v_j} = \sigma(f\Big(h^{(k)}_{v_i}\Big) \cdot f\Big(h^{(k)}_{v_j}\Big))\]

\noindent where $f(\cdot)$ is an NN module, and $\hat{y}_{v_i; v_j}$ represents the predicted probability of a relationship between nodes $v_i$ and $v_j$.

\section{Case Study}\label{sec: case study}

\subsection{Data Description}\label{sub: data description}

The dataset used to evaluate the proposed method is the Marklines automotive dataset, provided by a Japanese supplier information company \citep{fessina2024pattern}. 
This dataset, summarised in Table~\ref{tab: markline data}, includes five entity types: \texttt{company}, \texttt{customer}, \texttt{country}, \texttt{certificate}, and \texttt{product}. 
The \texttt{company} and \texttt{customer} entities represent the legal names of organisations (e.g., Hamenz For German Tech. Ind. (SAE)), while the \texttt{country} entity indicates the geographic location of a company (e.g., Egypt). 
The \texttt{certificate} entity refers to a company’s qualifications in the relevant sector, such as the International Organisation for Standardisation (ISO) certification (e.g., ISO9001), which ensures compliance with specific standards in areas like quality management, environmental management, or information security \citep{singels2001iso}. 
Finally, the \texttt{product} entity (e.g., Piston Ring Machining) describes the capability or category of a product, rather than its legal name.

The original Marklines dataset contains 41,693 unique companies, 2,353 customers, 79 countries, 5 certificates, and 927 products. 
After cleaning the dataset to remove missing and duplicate entries, the data was segmented by country to create subdatasets for evaluating the proposed approach.
Countries with insufficient data to construct a proper knowledge graph were excluded, based on the criteria that each subdataset must include all entity and relationship types and have at least 500 links of each type. 
Following this filtering process, data from ten countries were selected for model training and evaluation, as detailed in Table~\ref{tab: data description for each country.}.

The size and structure of the knowledge graphs constructed from these datasets vary significantly between countries, reflecting differences in dataset size, entity distribution, and relationship types. 
For example, the knowledge graph built from the China dataset includes 7,287 unique companies, 963 customers, 868 products, and 5 certificates, while constructed from the UK dataset, consists of 807 companies, 600 customers, 742 products and 5 certificates. 
These variations contribute to the heterogeneity of the knowledge graphs, showing additional challenges for model training. 
The differences in entity distributions and link types across countries further complicate the task of creating an FL framework capable of accommodating such diverse data structures.

\begin{table*}[th!]
\centering
\small
    \begin{tabular}{|p{1.2cm}|p{5.3cm}|p{2.2cm}|p{7.0cm}|} 
    \hline
     \textbf{Entity} & \textbf{Example} & \textbf{Unique Number} & \textbf{Notes} \\
    \hline
        Company & Hamenz For German Tech. Ind. (S.A.E.) & 41,693 & Companies who sell products. \\
        Customer & KOSHIN SEIKOSHO LTD.  & 2353 & Companies who buy products. \\
        Country & Egypt & 79 & \\
        Certificate & IS09001, QS9000, ISO/TS16949 & 5 & A company may have more than one certificate.\\
        Product & Piston Ring Machining & 927 & This is not the product name but the product description or capability. \\
     \hline
    \end{tabular}
\captionof{table}{Markline data (Entities and their unique numbers)}
\label{tab: markline data}
\end{table*}

\begin{figure*}[th!]
    \centering    
    \includegraphics[width=0.85\textwidth]{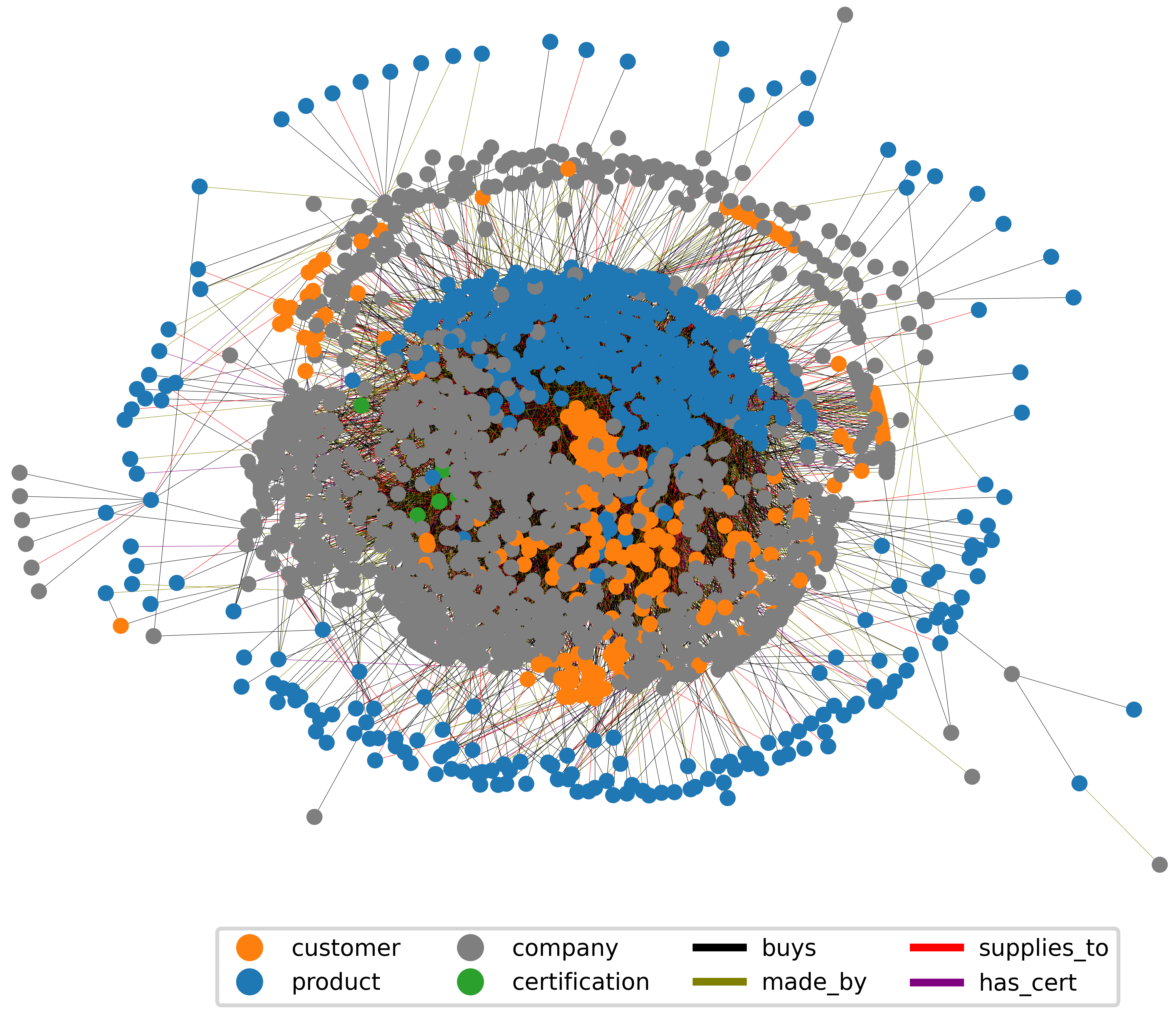}
    \caption{The supply chain knowledge graph constructed by the UK dataset.}
    \label{fig: (company, supplies-to, customer) with kg.}
\end{figure*}

\begin{table*}[th!]
\centering
\small
    \begin{tabular}{|p{1.45cm}|p{1.1cm}|p{1.1cm}|p{0.85cm}|p{1.2cm}|p{1.9cm}|p{1.9cm}|p{1.9cm}|p{1.7cm}|p{0.85cm}|} 
    \hline
     \textbf{Country Name} & \textbf{{Num-Company}} & \textbf{{Num-Customer}} & \textbf{Num-Product} & \textbf{Num-Certificate} & \textbf{Num-Edge}(Rate) (supplies\_to) & \textbf{Num-Edge}(Rate) (buys) & \textbf{Num-Edge}(Rate) (made\_by) & \textbf{Num-Edge}(Rate) (has\_cert) & \textbf{Total-Edges} \\
    \hline
    BARZIL & 173 & 200 & 341 & 5 & 1,057(9.22\%) & 5,163(45.03\%) & 4,742(41.36\%) & 504(4.40\%) & 11,466 \\
    CHINA & 7,287 & 963 & 868 & 5 & 33,706(17.64\%) & 79,209(41.46\%) & 65,422(34.25\%) & 12,704(6.65\%) & 191,041 \\
    GERMANY & 707 & 613 & 728 & 5 & 7,655(10.70\%) & 51,445(71.94\%) & 10,537(14.73\%) & 1,877(2.62\%) & 71,514 \\
    INDIA & 731 & 510 & 622 & 5 & 5,193(12.21\%) & 27,272(64.14\%) & 8,625(20.29\%) & 1,427(3.36\%) & 42,517 \\
    JAPAN & 2,975 & 980 & 883 & 5 & 16,227(13.30\%) & 71,582(58.66\%) & 30,559(25.04\%) & 3,671(3.01\%) & 122,039 \\
    KOREA & 710 & 291 & 685 & 5 & 2,573(10.94\%) & 12,409(52.78\%) & 6,816(28.99\%) & 1,715(7.29\%) & 23,513 \\
    TAIWAN & 186 & 239 & 374 & 5 & 957(7.56\%) & 6,189(48.87\%) & 4,490(35.45\%) & 1,029(8.13\%) & 12,665 \\
    THAILAND & 555 & 336 & 479 & 4 & 2,431(11.99\%) & 9,001(44.39\%) & 7,821(38.57\%) & 1,026(5.06\%) & 20,279 \\
    UK & 386 & 220 & 433 & 5 & 1,823(12.50\%) & 7,630(52.30\%) & 4,114(28.20\%) & 1,022(7.01\%) & 14,589 \\
    USA & 807 & 600 & 742 & 5 & 5,157(7.77\%) & 41,876(63.09\%) & 16,873(25.42\%) & 2,468(3.72\%) & 66,374 \\
    \hline
    \end{tabular}
\captionof{table}{Statistic information of the knowledge graph for each country.}
\label{tab: data description for each country.}
\end{table*}


\begin{table}[th!]
\centering
\small
    \begin{tabular}{|p{4.5cm}|p{2.5cm}|} 
    \hline
     Triple Name & Number of Links \\
    \hline
    (company, supplies\_to, customer) & 76,844 \\
    (customer, buys, product) & 737,703 \\
    (product, made\_by, company) & 160,170 \\
    (company, has\_cert, certification) & 27,478 \\
     \hline
    \end{tabular}
\captionof{table}{Ontologies used to construct knowledge graphs.}
\label{tab: reltionships}
\end{table}

\subsection{Experimental Settings} \label{sub: experimental settings}

For each country, we constructed a knowledge graph following Section~\ref{subsec: definition of SCKG}. 
The triplets used to build the knowledge graph were then partitioned into three sets: 70\% for training, 10\% for validation, and 20\% for testing. 
To enable the model to determine the existence of an edge, negative triplets were sampled in equal proportion to the positive triplets for training.

For the training process, the Adam optimizer \citep{kingma2014adam} was selected, and \textit{binary cross-entropy loss} was employed as the loss function, given that link prediction in the supply chain knowledge graph is formulated as a binary classification problem.

To evaluate the proposed approach, we compared it against multiple benchmarks: \textit{Local}, \textit{FLavg}, \textit{FLavgFT}, \textit{AdapFLavg}, and \textit{AdapFLavgFT} models. 
The \textit{Local} model benchmark involves training a separate GCN model for each country to predict relationships within its supply chain knowledge graph. 
The \textit{FLavg} model allows all countries to train their GCN and NN modules and then only share the parameters of the NN module to generate a global NN module through parameter averaging, enabling cross-country information sharing. 
\textit{FLavgFT} model further refines the \textit{FLavg} model by fine-tuning the last layer of the NN module to better adapt to local data.
The \textit{AdapFLavg} model \citep{zheng2024adaptive} introduces group-based FL, allowing countries with similar data patterns to collaborate in training a federated model. 
Initially, each country trains a local model and shares the parameters of its NN module with other countries for evaluation. Countries that demonstrate high performance when using each other’s parameters are identified as having similar patterns in their knowledge graphs and subsequently form a group to co-develop a federated model.
\textit{AdapFLavgFT}, similar to \textit{FLavgFT}, fine-tunes the last layer of the NN module in \textit{AdapFLavg} model to enhance local adaptation. 
To ensure reliable results, each experiment was conducted 20 times, with model parameters reinitialised for every iteration.

To measure performance, two metrics were employed: the Receiver Operating Characteristic Area Under the Curve (ROC-AUC) score and Average Precision (AP). 
ROC-AUC combines the Receiver Operating Characteristic (ROC) curve with the Area Under the Curve (AUC) to evaluate the model's ability to distinguish between two classes, with scores ranging from 0 to 1. 
Higher scores indicate better classification performance. 
Similarly, AP, ranging from 0 to 1, evaluates the model’s accuracy in predicting true supply chain relationships, where higher values stand for better performance.

In addition, we used the Friedman test \citep{sheldon1996use} to analyse performance differences across the five models and the Post-hoc test \citep{ruxton2008time} to identify specific model pairs with significant differences. 
The Friedman test assesses whether significant differences exist among related models based on ranked data but does not pinpoint where those differences occur. 
The Post-hoc test complements the Friedman test by identifying specific pairs of models with statistically significant performance differences, serving as a secondary analysis.

\subsection{Experimental Results} \label{sub: results}

Table~\ref{tab: AUC results.} and Table~\ref{tab: AP results.} present the ROC-AUC and AP scores achieved by the five models in predicting four types of relationships within the supply chain knowledge graphs from ten countries. 
The five models are categorised into three groups: 1) \textit{LocalM}, 2) \textit{FLavg} and \textit{FLavgFT}, and 3) \textit{AdapFLavg} and \textit{AdapFLavgFT}. 
For clarity, these groups are referred to as 1) the local model, 2) FL-based models, and 3) AdapFL-based models in the subsequent discussion.
The results allow us to draw several key observations, which are detailed below.

\begin{table*}[th!]
\centering
\small
\begin{tabular}{|p{2.0cm}|p{3.5cm}|p{3.5cm}|p{3.5cm}|p{3.5cm}|} 
\hline
\multirow{2}{*}{\textbf{Country Name}} & \textbf{\textit{supplier\_to}} & \textbf{\textit{buys}} & \textbf{\textit{made\_by}} & \textbf{\textit{has\_cert}} \\
\cline{2-5}
 & LocalM/FLavg/FLavgFT/ AdapFLavg/AdapFLavgFT & LocalM/FLavg/FLavgFT/ AdapFLavg/AdapFLavgFT & LocalM/FLavg/FLavgFT/ AdapFLavg/AdapFLavgFT & LocalM/FLavg/FLavgFT/ AdapFLavg/AdapFLavgFT \\
\hline
BARZIL  & 0.7621/0.7808/0.7636/ 0.7938/\textbf{0.8427} & 0.8296/0.7752/0.7598/ 0.8253/\textbf{0.8482} & 
          0.7009/0.7013/\textbf{0.7228}/ 0.7102/0.7182 & 0.6534/0.7486/0.7479/ \textbf{0.8604}/0.8576 \\
CHINA   & 0.8226/0.7980/0.7884/ 0.8436/\textbf{0.8556} & 0.8711/0.7925/0.7959/ 0.8817/\textbf{0.8976} & 
          0.8043/0.7696/0.7785/ 0.8180/\textbf{0.8436} & 0.7121/\textbf{0.7488}/0.7372/ 0.7461/0.7436 \\
GERMANY & 0.8500/0.8261/0.8344/ \textbf{0.9196}/0.9085 & 0.8632/0.7897/0.7858/ \textbf{0.8668}/0.8518 &            
          0.7567/0.7292/0.7222/ 0.7760/\textbf{0.7921} & 0.5687/0.7000/0.6496/ 0.6928/\textbf{0.7001}  \\
INDIA   & 0.7537/0.7655/0.7650/ \textbf{0.8383}/0.8107 & \textbf{0.8352}/0.7521/0.7520/ 0.8236/0.8171 &            
          0.7380/0.7274/0.7355/ 0.7360/\textbf{0.7539} & 0.6003/0.6948/0.6281/ \textbf{0.7071}/0.6757  \\
JAPAN   & 0.8056/0.7789/0.7867/ \textbf{0.8826}/0.8670 & 0.8394/0.7594/0.7689/ 0.8334/\textbf{0.8777} &            
          \textbf{0.8237}/0.7804/0.7800/ 0.8072/0.8118 & 0.8113/0.8639/0.8412/ 0.86391/\textbf{0.8770}  \\
KOREA   & 0.8415/0.8505/\textbf{0.8571}/ 0.8319/0.8381 & 0.8374/0.8028/0.7981/ 0.8418/\textbf{0.8602} & 
          0.7125/0.6985/0.7000/ 0.6978/\textbf{0.7503} & 0.6700/0.7254/0.6968/ \textbf{0.7441}/0.7317 \\
TAIWAN  & 0.7485/0.7484/0.7397/ 0.7931/\textbf{0.8399} & 0.8105/0.7565/0.7518/ 0.8014/\textbf{0.8424} & 
          0.7046/0.6854/0.7050/ \textbf{0.7069}/0.6916 & 0.6421/0.6704/0.6716/ 0.7182/\textbf{0.7303}  \\
THAILAND& 0.7368/0.7773/0.7917/ 0.8642/\textbf{0.8352} & \textbf{0.8331}/0.7707/0.7769/ 0.8085/0.8305 & 
          0.7411/0.7585/0.7538/ \textbf{0.8059}/0.7898 & 0.5357/0.6641/0.5774/ \textbf{0.7624}/0.7111  \\
USA     & 0.8132/0.7895/0.8106/ \textbf{0.8809}/0.8169 & 0.8596/0.8074/0.8026/ \textbf{0.8808}/0.8373 & 
          0.7672/0.7419/0.7514/ 0.7925/\textbf{0.7965} & 0.6101/0.7103/0.6530/ 0.6836/\textbf{0.7391}  \\
UK      & 0.7328/0.7964/0.7943/ \textbf{0.8066}/0.8007 & 0.8493/0.7928/0.7984/ 0.8663/\textbf{0.8846} & 
          0.7154/0.7146/0.7197/ 0.7262/\textbf{0.7556} & 0.6075/0.6878/0.7409/ \textbf{0.7887}/0.7294 \\
\hline
\end{tabular}
\captionof{table}{ROC-AUC scores achieved by all five models for predicting four types of relationships in supply chain knowledge graphs across ten countries.}
\label{tab: AUC results.}
\end{table*}

\begin{table*}[th!]
\centering
\small
\begin{tabular}{|p{2.0cm}|p{3.5cm}|p{3.5cm}|p{3.5cm}|p{3.5cm}|} 
\hline
\multirow{2}{*}{\textbf{Country Name}} & \textbf{\textit{supplier\_to}} & \textbf{\textit{buys}} & \textbf{\textit{made\_by}} & \textbf{\textit{has\_cert}} \\
\cline{2-5}
 & LocalM/FLavg/FLavgFT/ AdapFLavg/AdapFLavgFT & LocalM/FLavg/FLavgFT/ AdapFLavg/AdapFLavgFT & LocalM/FLavg/FLavgFT/ AdapFLavg/AdapFLavgFT & LocalM/FLavg/FLavgFT/ AdapFLavg/AdapFLavgFT \\
\hline
BARZIL  & 0.7687/0.8087/0.7817/ 0.8159/\textbf{0.8242} & 0.8280/0.7831/0.7696/ 0.8117/\textbf{0.8404} & 0.7388/0.7569/\textbf{0.7665}/ 0.7608/0.7608 & 0.6434/0.6976/0.6957/ \textbf{0.8583}/0.8424 \\
CHINA   & 0.8218/0.8208/0.8117/ \textbf{0.8489}/0.8335 & 0.8728/0.7978/0.8035/ 0.8742/\textbf{0.8967} & 0.8256/0.8098/0.8160/ 0.8387/\textbf{0.8449} & 0.6746/\textbf{0.7104}/0.6952/ 0.7101/0.7049 \\
GERMANY & 0.8550/0.8457/0.8525/ \textbf{0.9218}/0.9090 & \textbf{0.8659}/0.8074/0.8022/ 0.8542/0.8575 & 0.7847/0.7738/0.7667/ 0.7725/\textbf{0.7903} & 0.5703/0.6555/0.6187/ 0.6609/\textbf{0.6996} \\
INDIA   & 0.7684/0.7923/0.8022/ \textbf{0.8268}/0.7927 & \textbf{0.8371}/0.7697/0.7646/ 0.8274/0.8217 & 0.7669/0.7620/\textbf{0.7777}/ 0.7755/0.7621 & 0.5927/0.6506/0.5961/ \textbf{0.6916}/0.6428 \\
JAPAN   & 0.8209/0.8162/0.8283/ \textbf{0.8927}/0.8751 & 0.8503/0.7853/0.7967/ 0.8446/\textbf{0.8748} & \textbf{0.8400}/0.8180/0.8166/ 0.8357/0.8360 & 0.7824/0.8365/0.8055/ 0.8486/\textbf{0.8604} \\
KOREA   & 0.8625/0.8790/\textbf{0.8939}/ 0.8752/0.8619 & 0.8497/0.8292/0.8190/ 0.8434/\textbf{0.8551} & 0.7495/0.7505/0.7487/ 0.7523/\textbf{0.7645} & 0.6287/0.6780/0.6460/ \textbf{0.7249}/0.6945 \\
TAIWAN  & 0.7566/0.7682/0.7549/ 0.8100/\textbf{0.8552} & 0.8101/0.7568/0.7450/ 0.8049/\textbf{0.8162} & 0.7444/0.7275/0.7498/ \textbf{0.7544}/0.7446 & 0.6147/0.6493/0.6352/ \textbf{0.7014}/0.6676 \\
THAILAND& 0.7477/0.8024/0.8185/ \textbf{0.8717}/0.8435 & \textbf{0.8389}/0.7794/0.7896/ 0.8115/0.8359 & 0.7729/0.7978/0.7975/ \textbf{0.8276}/0.8095 & 0.5527/0.6378/0.5887/ \textbf{0.7377}/0.6788 \\
USA     & 0.8250/0.8164/0.8422/ \textbf{0.8678}/0.8434 & 0.8620/0.8175/0.8147/ \textbf{0.8861}/0.8476 & 0.7926/0.7838/0.7897/ \textbf{0.8026}/0.8004 & 0.5960/0.6616/0.6095/ 0.6421/\textbf{0.7037} \\
UK      & 0.7398/\textbf{0.8198}/0.8051/ 0.8110/0.8049 & 0.8480/0.7956/0.8029/ 0.8536/\textbf{0.8781} & 0.7540/0.7536/0.7666/ 0.7643/\textbf{0.7885} & 0.6105/0.6383/0.6814/ \textbf{0.7582}/0.6580 \\
\hline
\end{tabular}
\captionof{table}{AP scores achieved by all five models for predicting four types of relationships in supply chain knowledge graphs across ten countries.}
\label{tab: AP results.}
\end{table*}

\subsubsection{Comparison of the Five Models}

For both ROC-AUC and AP metrics, AdapFL-based models demonstrate superior performance compared to local and FL-based models in nearly all countries, particularly in predicting \textit{has\_cert} relationships. 
These relationships are the most challenging to predict among the four types due to their relatively small representation in the datasets. 
As shown in Table~\ref{tab: data description for each country.}, the \textit{has\_cert} edges constitute less than 5\% of the total edges in most countries, and the number of certifications is limited to five in almost all cases far fewer than other entity types. 
Despite this imbalance, AdapFL-based models outperform other models in predicting \textit{has\_cert} relationships. 
This success can be attributed to their ability to share relevant and useful information in clusters of countries with similar data patterns, rather than share all the accessible information when using FL-based models.

For countries with larger knowledge graphs, such as CHINA, GERMANY, JAPAN, KOREA, and USA, the local model exhibits better performance in predicting \textit{supplies\_to}, \textit{buys}, and \textit{made\_by} relationships compared to FL-based models. 
Larger datasets enable the local model to learn effectively without relying on shared information from other countries. 
In these cases, shared information from other countries may introduce noise, which diminishes prediction accuracy. 
For example, CHINA and JAPAN have the largest supply chain knowledge graphs, enabling the local model to excel in predicting three types of relationships excluding the relationship of \textit{has\_cert}. 
These findings align with the results of \citep{zheng2023federated}, suggesting that FL-based models are better suited for smaller datasets rather than large ones.

\begin{figure*}[htbp]
    \centering
    \newcolumntype{C}[1]{>{\centering\arraybackslash}m{#1}}
    \begin{tabular}{C{0.27\textwidth} C{0.20\textwidth} C{0.20\textwidth} C{0.24\textwidth}}
        \textbf{\textit{supplies\_to}} & \textbf{\textit{buys}} & \textbf{\textit{made\_by}} & \textbf{\textit{has\_cert}} \\
        \begin{subfigure}
            \centering
            \includegraphics[width=\linewidth]{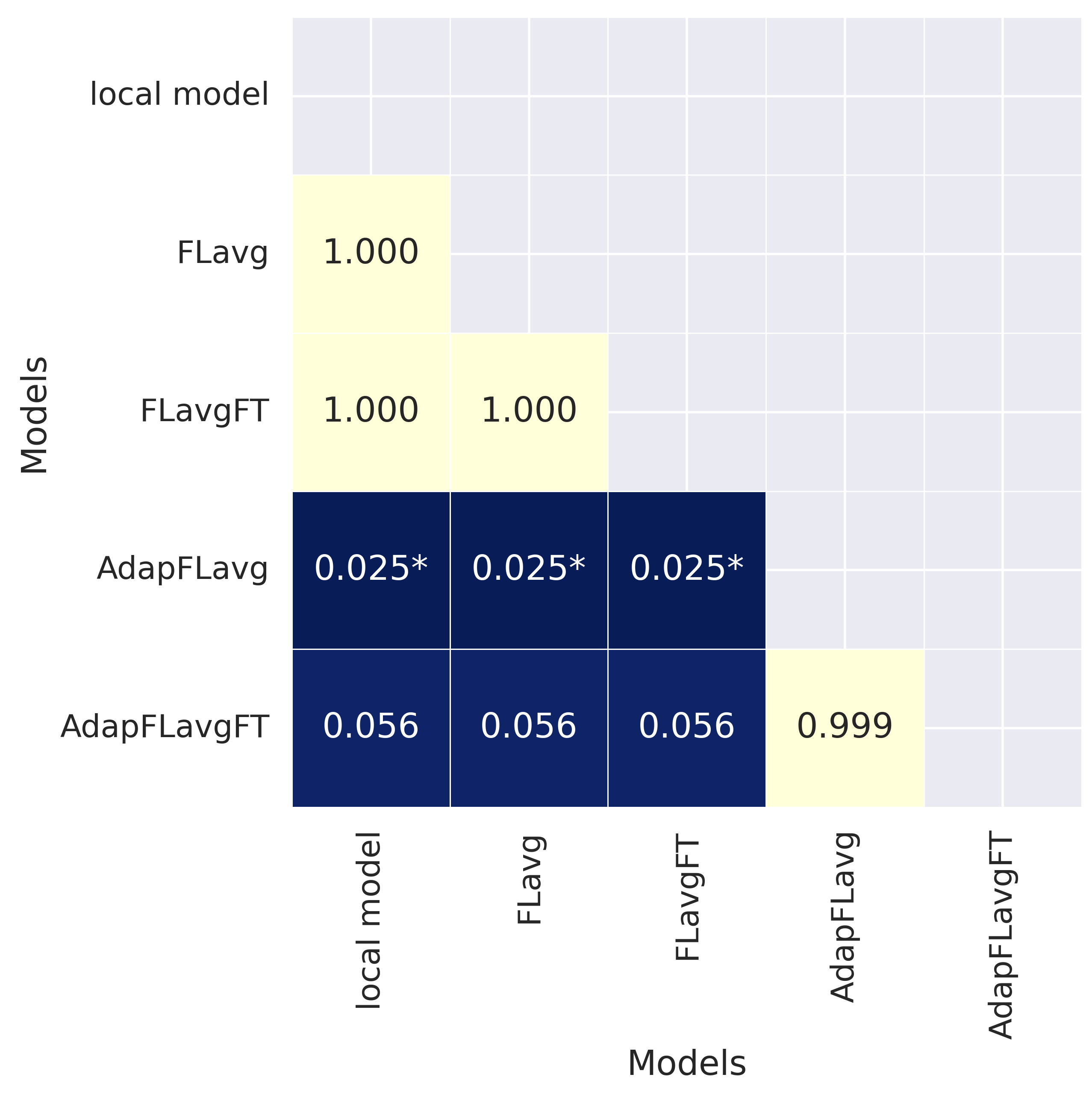}
        \end{subfigure}
        &
        \begin{subfigure}
            \centering
            \includegraphics[width=\linewidth]{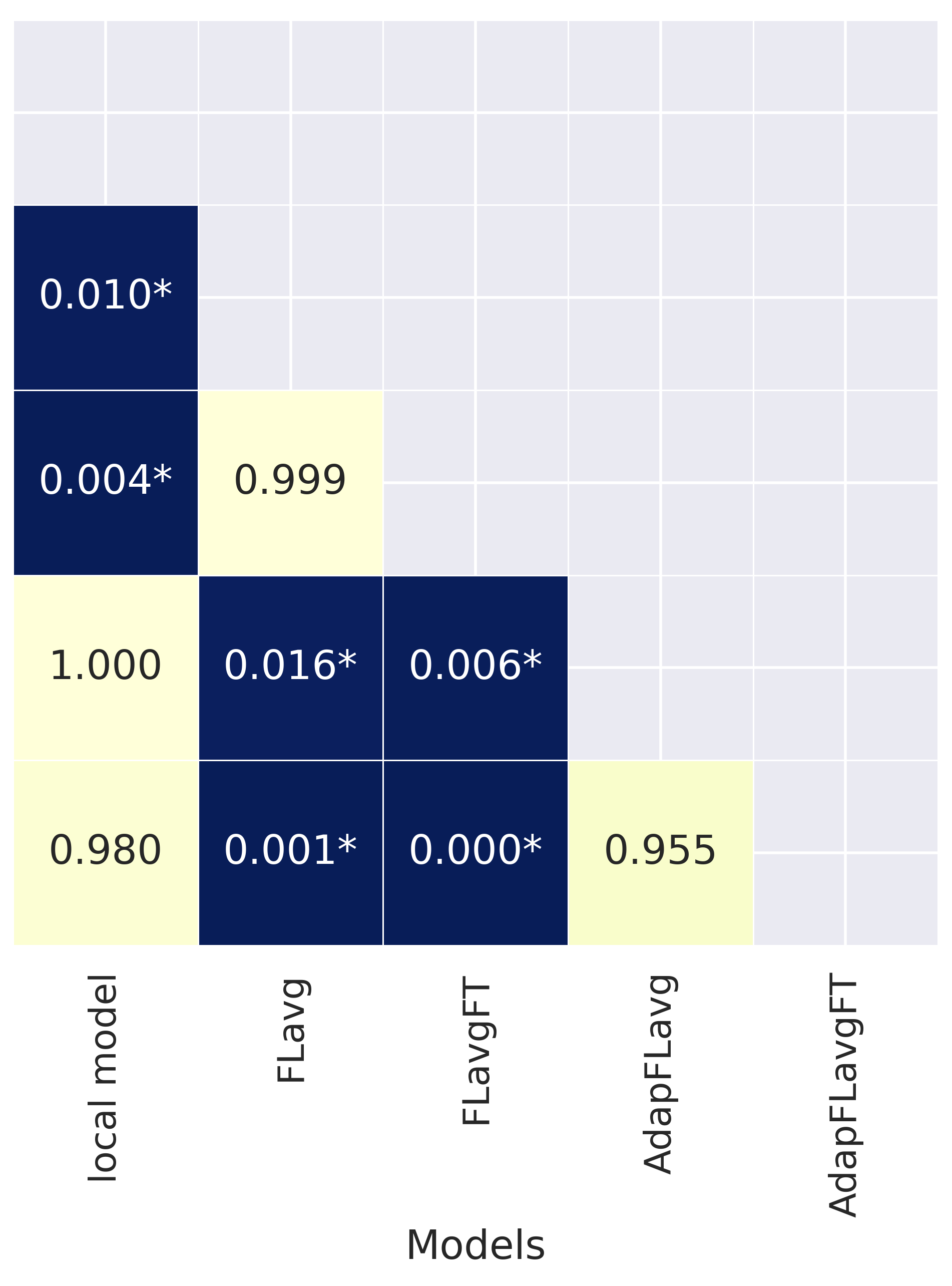}
        \end{subfigure}
        &
        \begin{subfigure}
            \centering
            \includegraphics[width=\linewidth]{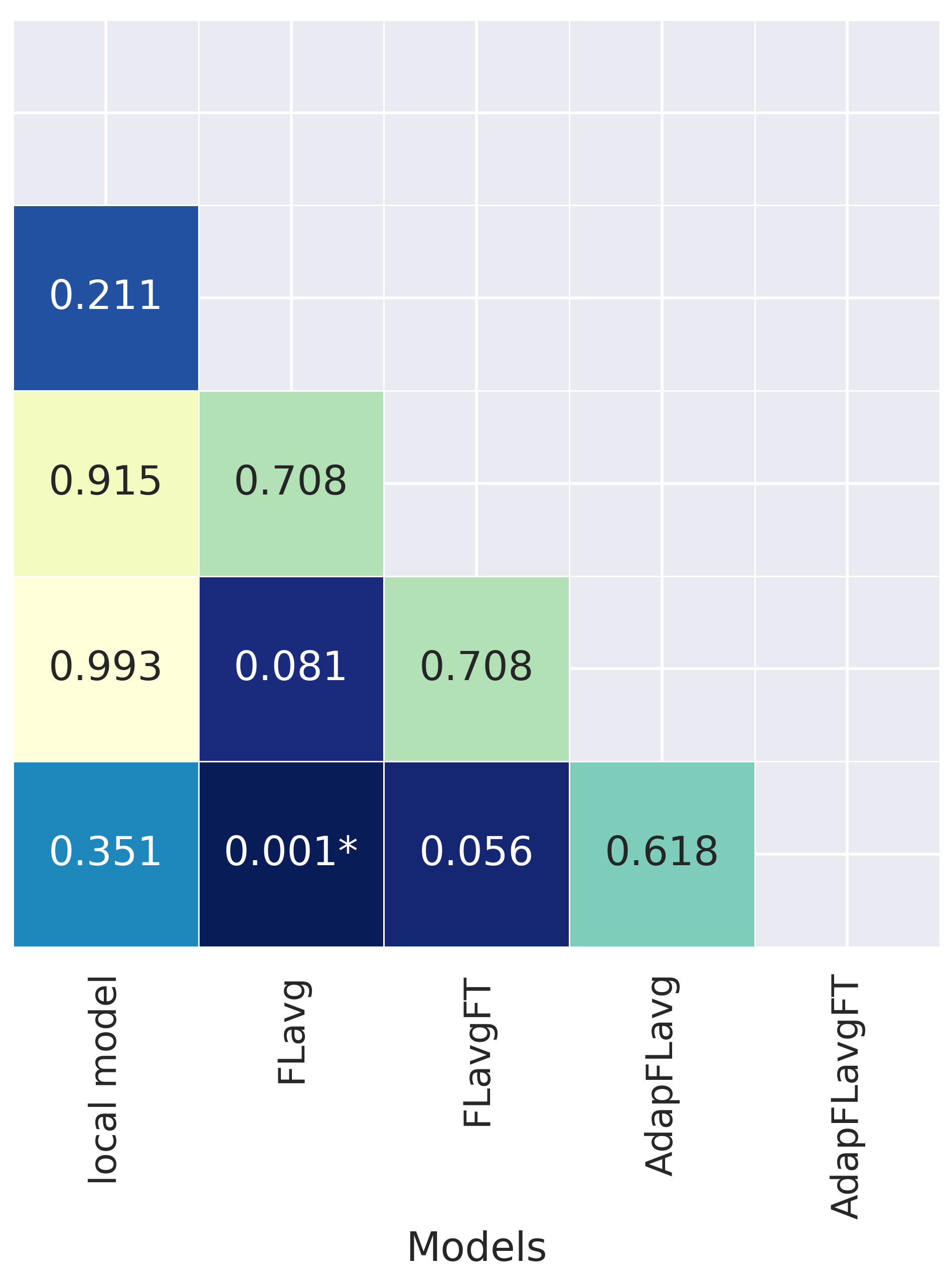}
        \end{subfigure}
        &
        \begin{subfigure}
            \centering
            \includegraphics[width=\linewidth]{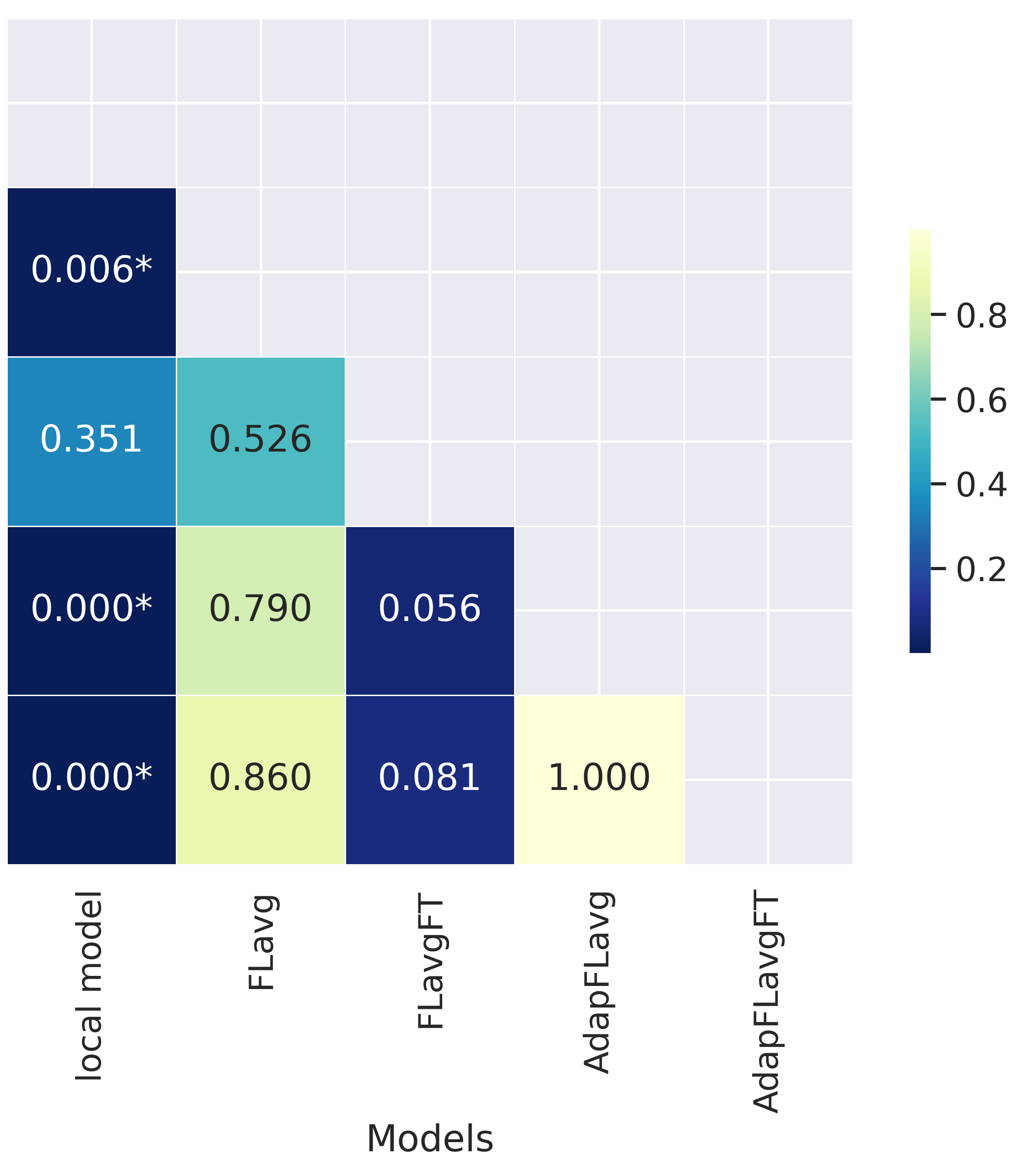}
        \end{subfigure}
        \\
        \begin{subfigure}
            \centering
            \includegraphics[width=\linewidth]{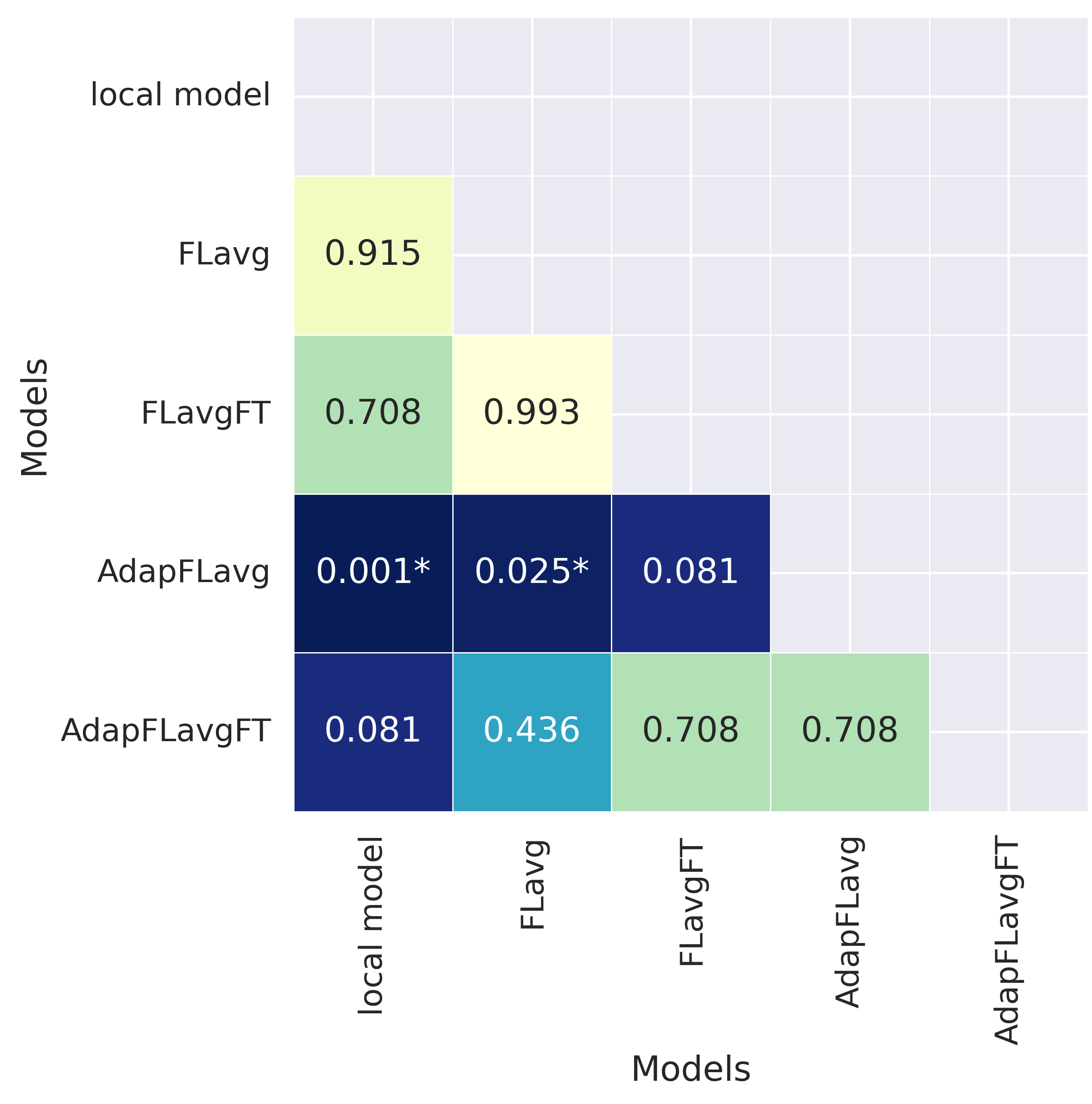}
        \end{subfigure}
        &
        \begin{subfigure}
            \centering
            \includegraphics[width=\linewidth]{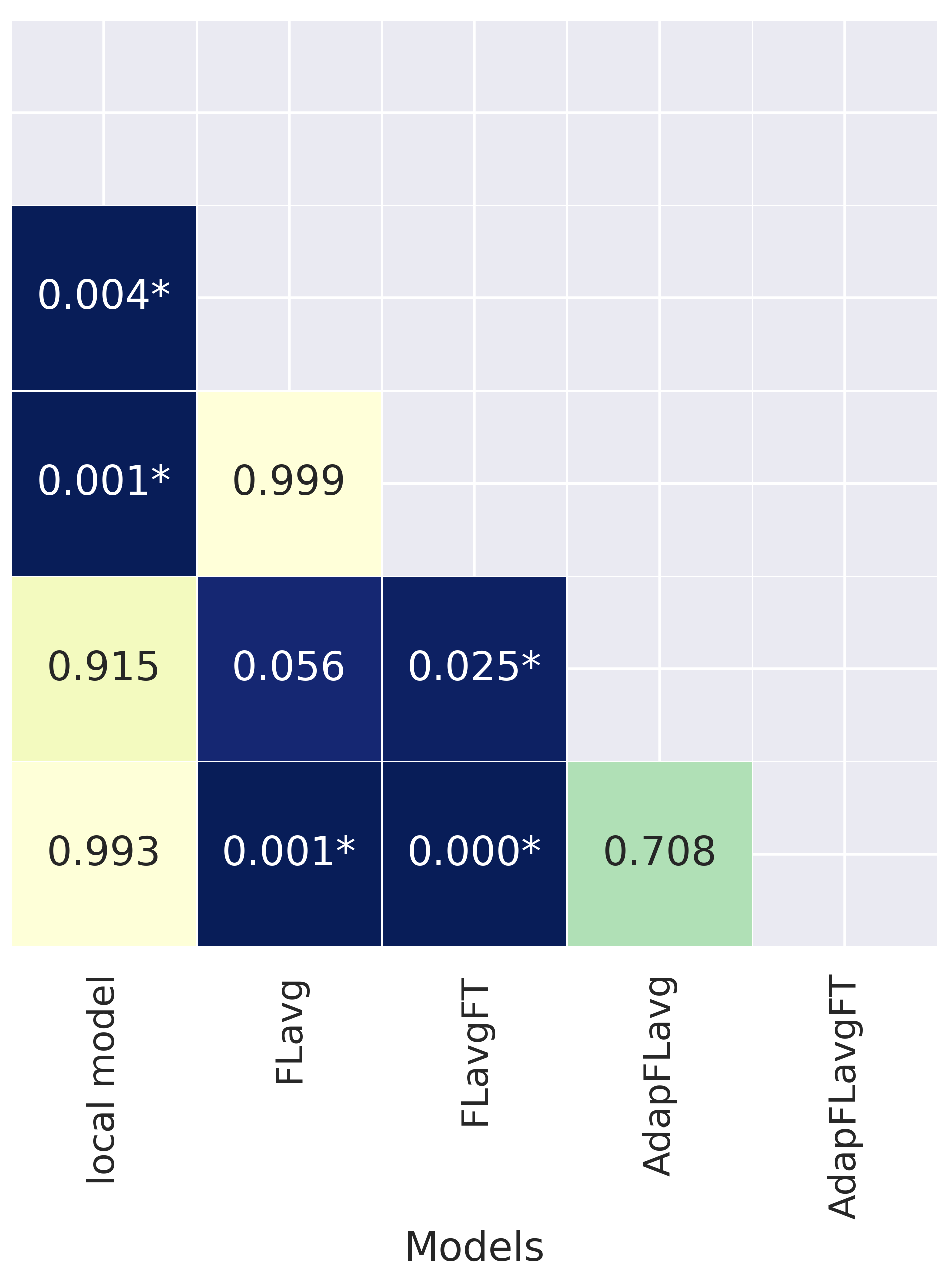}
        \end{subfigure}
        &
        \begin{subfigure}
            \centering
            \includegraphics[width=\linewidth]{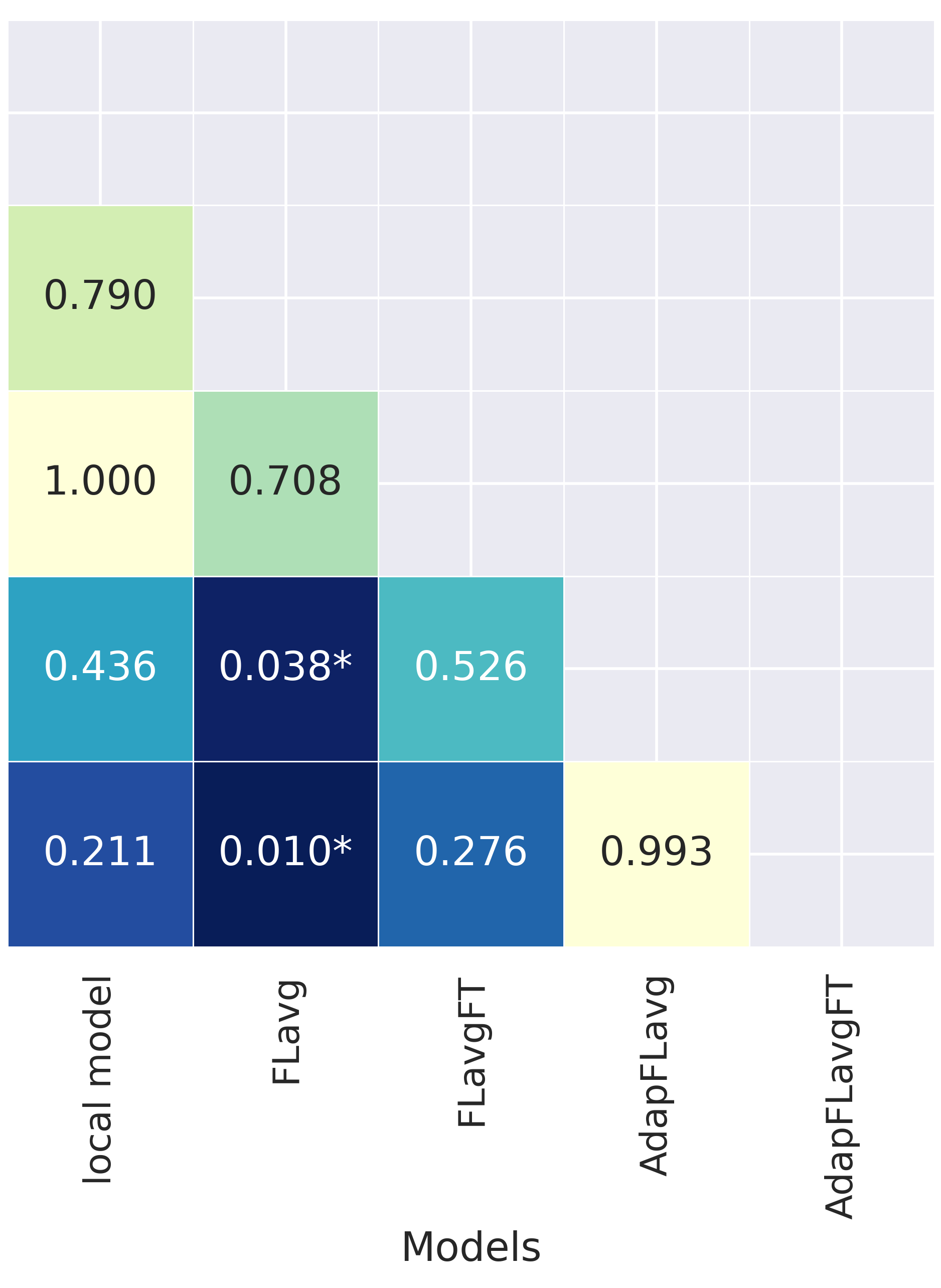}
        \end{subfigure}
        &
        \begin{subfigure}
            \centering
            \includegraphics[width=\linewidth]{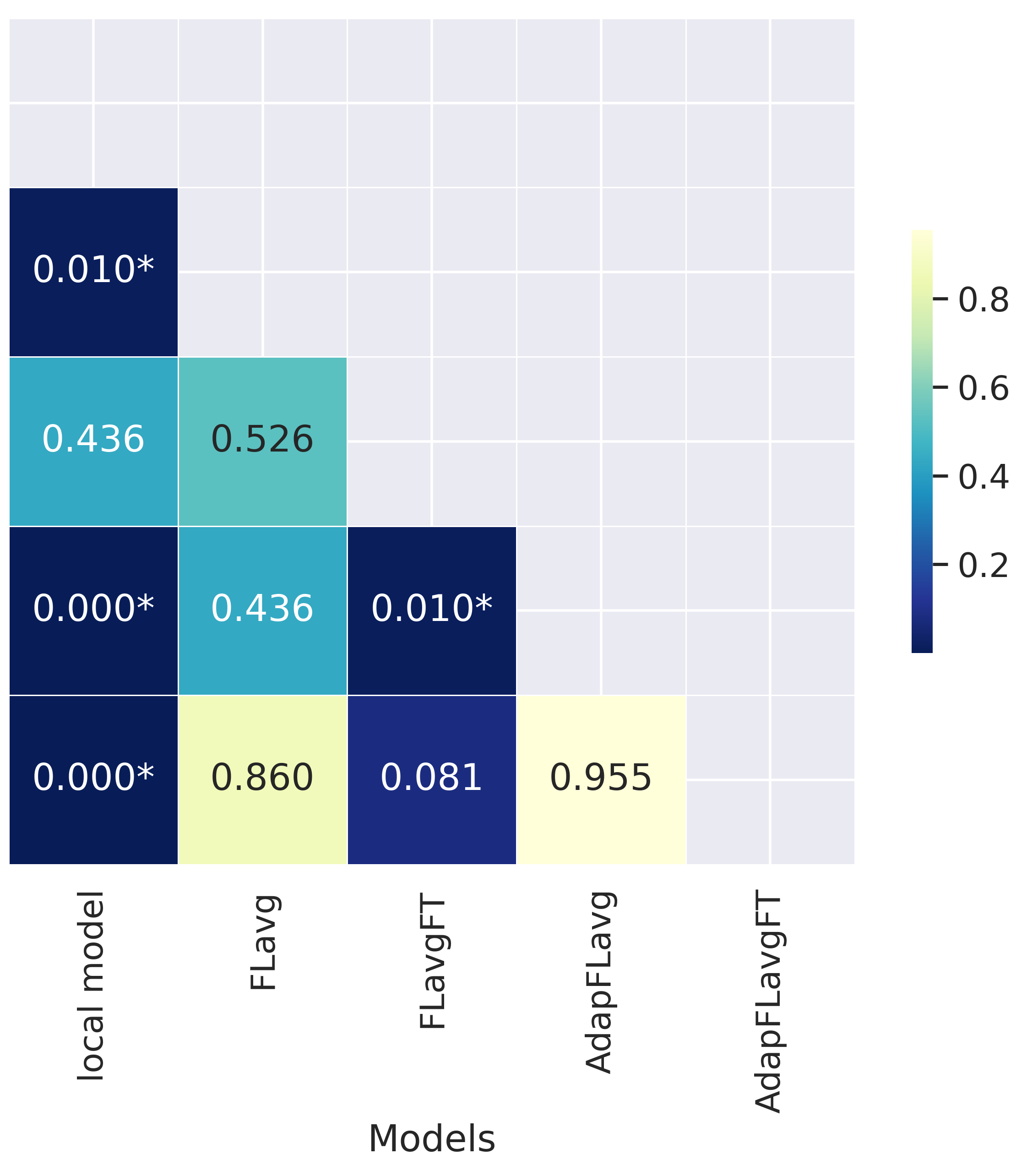}
        \end{subfigure}
        \\
    \end{tabular}
    
    \caption{The results of the significance test highlight the differences among the five models.}
    \label{fig: significant test.}
\end{figure*}

To understand the significance of performance differences among the five models, we employed the Friedman test and visualised the results using the Post-hoc test in Figure~\ref{fig: significant test.}, with a standard p-value of 0.05 used to indicate statistical significance. 
The Friedman test evaluates overall performance differences, while the Post-hoc test identifies specific pairs of models with significant differences. 
The first row of results in Figure~\ref{fig: significant test.} is based on ROC-AUC, while the second row is derived from AP, reflecting slight differences between the metrics. 
ROC-AUC measures the ability to distinguish between relationships and non-relationships, whereas AP focuses on identifying true relationships.

The most notable finding is that AdapFL-based models significantly improve the prediction of \textit{has\_cert} relationships compared to the local model. 
This result aligns with observations from Table~\ref{tab: AUC results.}, Table~\ref{tab: AP results.}, and \citep{zheng2023federated}, reaffirming that FL-based models can provide substantial benefits for smaller datasets. 
In this case, AdapFL-based models effectively address the scarcity of \textit{has\_cert} relationships in country-level supply chain knowledge graphs. 
Furthermore, AdapFL-based models significantly enhance prediction accuracy compared to FL-based models for \textit{supplies\_to} and \textit{buys} relationships. 
Specifically, for \textit{buys} relationships, AdapFL-based and local models perform similarly and both significantly outperform FL-based models, as evidenced by Figure~\ref{fig: significant test.}, Table~\ref{tab: AUC results.}, and Table~\ref{tab: AP results.}.

\subsubsection{Comparison of Four Types of Relationships}

Among the four types of relationships, all five models perform worst in predicting the \textit{has\_cert} relationships for each country. This is primarily due to the low proportion of \textit{has\_cert} relationships, which constitute less than 5\% of all edges in most countries, as shown in Table~\ref{tab: data description for each country.}. 
The scarcity of \textit{has\_cert} relationships makes it challenging for the models to effectively learn patterns from these relationships among a large number of other types of relationships.

Oppositely, the five models achieve their best performance in predicting the \textit{buys} relationships. 
This is attributable to the \textit{buys} relationships having the largest proportion among the four types, which provides the models with sufficient data to learn robust patterns and distinguish these relationships from others, resulting in better prediction accuracy.

Interestingly, the models also perform better in predicting \textit{supplies\_to} relationships than \textit{made\_by} relationships, despite the latter being more numerous in each country’s supply chain knowledge graph. 
To understand this discrepancy, we analysed networks with isolated relationship types by calculating key structural metrics, including average degree, clustering coefficient, density, closeness, and betweenness, as shown in Tables~\ref{tab: supplier to network}, \ref{tab: buys network}, \ref{tab: made by network}, and \ref{tab: has cert network} in appendix. 
The \textit{supplies\_to} networks exhibit higher clustering coefficients, density, closeness, and betweenness compared to the \textit{made\_by} networks. 
These structural properties suggest that the patterns within \textit{supplies\_to} networks are easier for models to identify, leading to better performance in predicting \textit{supplies\_to} relationships compared to \textit{made\_by} relationships.

\begin{figure*}[htbp]
    \centering
    \begin{subfigure}[]
        \centering
        \includegraphics[width=9.0cm]{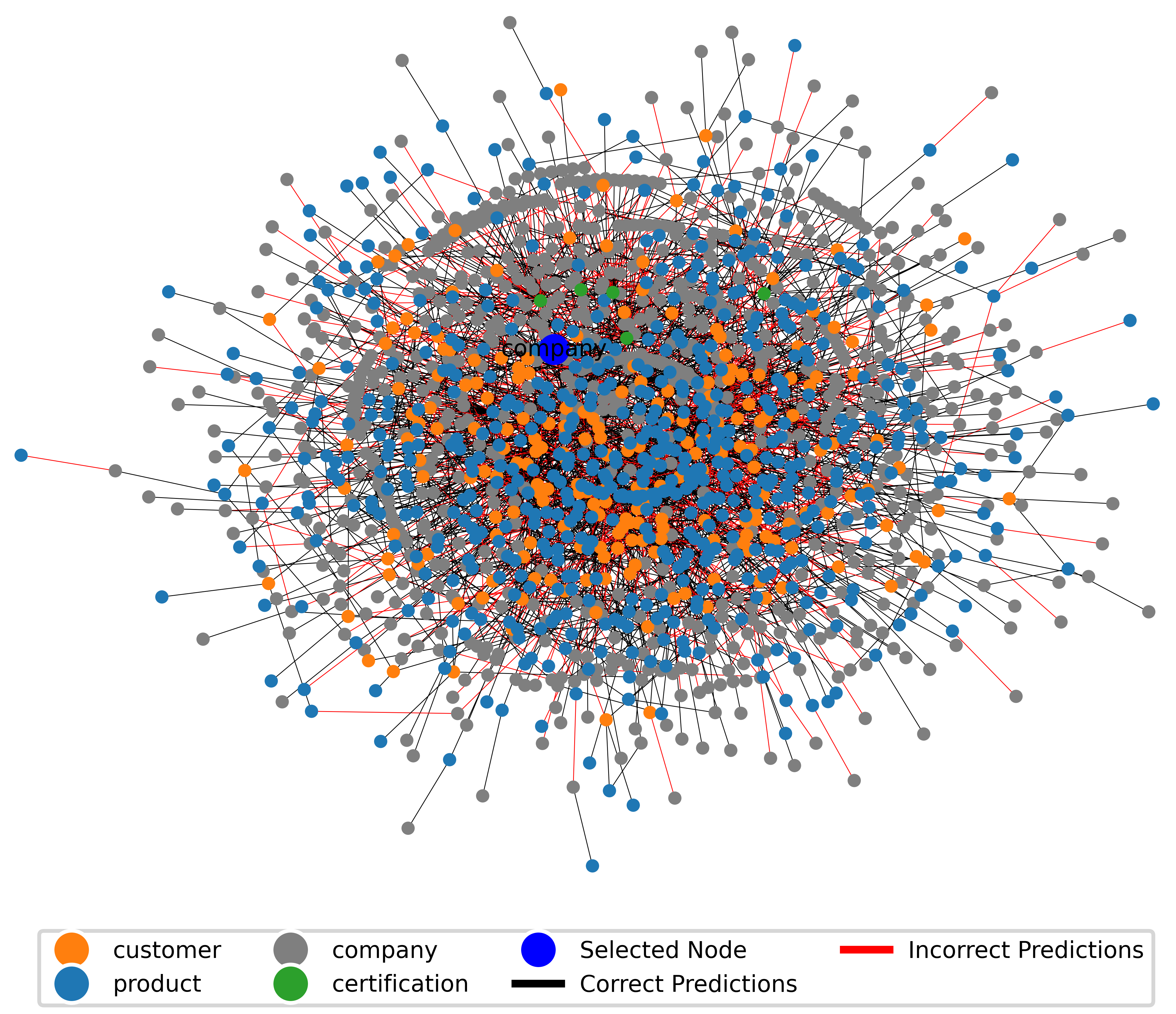}
    \end{subfigure}
    \hspace{0.5cm}
    \begin{subfigure}[]
        \centering
        \includegraphics[width=8.0cm]{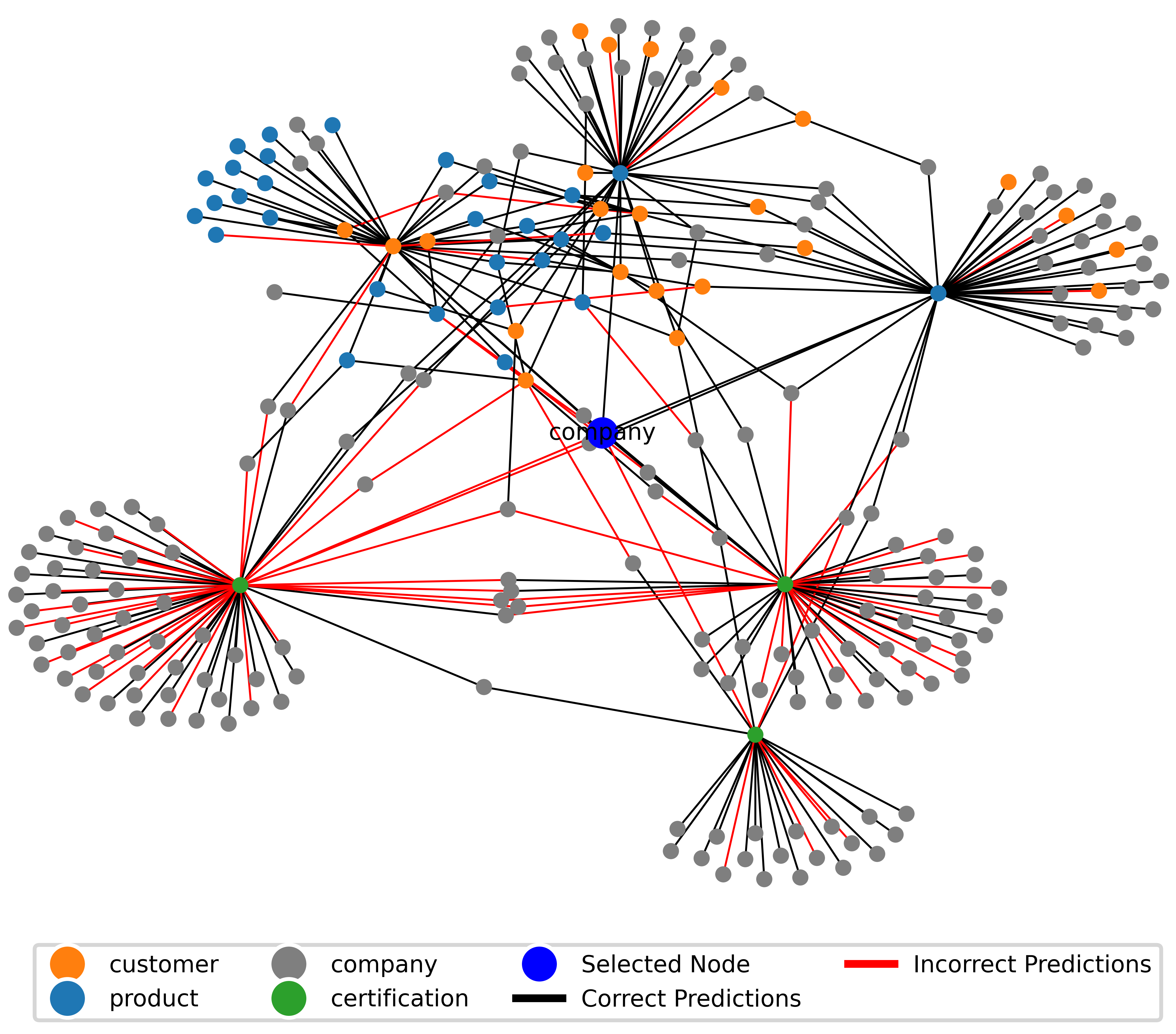}
    \end{subfigure}
    \caption{Predicted results on the UK supply chain knowledge graph.}
    \label{fig: predicted links}
\end{figure*}

Figure~\ref{fig: predicted links} illustrates the sub-knowledge graph of a real-world UK supply chain, along with predicted relationships generated by the AdapFLavg model. 
The sub-knowledge graph is constructed using the test dataset to evaluate the proposed approach. 
In Figure~\ref{fig: predicted links} (a), edges correctly predicted by the model are shown in black, while incorrect predictions are marked in red. 
Figure~\ref{fig: predicted links} (b) focuses on a randomly selected node and its two-hop neighbours, where the GCN module aggregates information for the selected node’s representation and relationship predictions.
The results highlight that predictions for edges around customers and products (e.g., \textit{supplies\_to} and \textit{buys}) are mostly accurate (black edges), whereas predictions for edges related to certifications (e.g., \textit{has\_cert}) are more likely to be incorrect. 
This observation aligns with the findings in Table~\ref{tab: AUC results.} and Table~\ref{tab: AP results.}, further showing the challenges in predicting \textit{has\_cert} relationships again.

\subsubsection{Comparison of Results among Ten Countries}

Among the ten countries, all five models achieve more accurate predictions across all four relationship types for CHINA, GERMANY, and JAPAN. 
This can be attributed to the larger size of the supply chain knowledge graphs in these countries compared to the other seven. 
This result aligns with a common observation in machine learning: ``larger datasets generally enable models to deliver better performance by providing more comprehensive patterns for learning'' \citep{halevy2009unreasonable}.

An interesting observation is that JAPAN is the only country where the average ROC-AUC and AP scores for all five models exceed 80\% across the four relationship types, despite its knowledge graph (122,039 edges) being smaller than CHINA’s (191,041 edges). 
A potential explanation lies in the lower diversity of company nodes in JAPAN’s graph (2,975 nodes) compared to CHINA’s (7,287 nodes). 
Higher node diversity often introduces greater heterogeneity, which can complicate the learning process and hinder the models’ ability to identify consistent patterns for relationship prediction.

Conversely, in countries with smaller supply chain knowledge graphs, such as BRAZIL, TAIWAN, and the UK, all models demonstrate lower prediction accuracy across almost all relationship types. 
This finding further demonstrates the importance of data volume in enhancing machine learning models’ ability to learn consistent patterns, leading to improved prediction accuracy.






\section{Conclusions and Managerial Implications} \label{sec: conclusion}

This study presents a novel approach to enhancing supply chain visibility by integrating Federated Learning (FL) and Graph Convolutional Networks (GCNs) to predict relationships within supply chain knowledge graphs. 
In today’s highly interconnected and complex global economy, disruptions such as the COVID-19 pandemic and geopolitical crises have underscored the vulnerabilities of supply chains and the critical need for improved visibility. 
Our methodology directly addresses key challenges, including privacy, security, and regulatory constraints, which often hinder traditional data-sharing approaches. 
By enabling collaborative learning across multiple parties without requiring raw data exchange, FL ensures compliance with privacy regulations and maintains data security. Simultaneously, GCNs capture intricate relational structures within knowledge graphs, uncovering hidden connections and enhancing predictive accuracy.

Unlike traditional firm- or product-level networks, our framework leverages the comprehensive structure of supply chain knowledge graphs, integrating diverse entities and relationships to model the complexity of real-world supply chains. 
This holistic approach facilitates the discovery of hidden relationships, such as compliance risks indicated by new \textit{has\_cert} connections or emerging market trends revealed through \textit{buys} relationships. 
These insights provide a deeper understanding of supply chain dynamics, enabling more effective risk management, operational optimisation, and strategic decision-making.

The proposed approach offers significant practical implications. 
Enhanced supply chain visibility empowers managers to proactively identify potential disruptions, mitigate risks, and explore new business opportunities. 
For example, by uncovering supplier dependencies or understanding purchasing patterns, organisations can better align their strategies with market dynamics and regulatory frameworks. 
Furthermore, the FL framework encourages cross-border collaboration among multinational corporations, enabling regions to share insights and co-develop solutions while preserving data privacy and respecting local regulations. 
With capabilities for scenario analysis, companies can evaluate the potential impacts of disruptions or policy changes, supporting more resilient and adaptive supply chain strategies.

Despite its strengths, this study also highlights limitations that suggest avenues for future research. 
Implementing FL across highly heterogeneous datasets within complex supply chain networks may lead to performance trade-offs and computational challenges. 
Furthermore, regional variations in privacy laws and regulatory requirements could complicate the broader adoption of this framework. 
Future work could focus on refining FL protocols to address data heterogeneity and improve computational efficiency, exploring advanced GCN architectures for enhanced link prediction accuracy, and incorporating real-time data streams to enable dynamic adaptability. 
These advancements hold promise for further enhancing supply chain visibility and resilience, ensuring that supply chains are better equipped to navigate the complexities of the modern global economy.

\section*{Data Availability Statement}

Due to the commercially sensitive nature of this research, supporting data is not available.

\section*{Acknowledgements}
This work was supported by the Engineering and Physical Sciences Research Council (grant number EP/W019868/1).

{\small
	\bibliographystyle{tfcad}
	\bibliography{reference.bib}
}

\section*{Appendix A}

\begin{table*}[ht]
\centering
\begin{filecontents*}{tables/customer_network_params_new.csv}
,country,numHead,numTail,numEdge,averageDegree,density,closeness,betweenness,averageClusteringCoeffi
9,BRAZIL,173,200,1057,5.6676,0.0152,0.2853,0.0067,0.0000
5,CHINA,7287,963,33706,8.1731,0.0010,0.2608,0.0003,0.0001
4,GERMANY,707,613,7655,11.6960,0.0089,0.3329,0.0016,0.0152
3,INDIA,731,510,5193,8.4715,0.0069,0.3012,0.0019,0.0077
8,JAPAN,2975,980,16227,8.3860,0.0022,0.2836,0.0007,0.0183
7,KOREA,710,291,2573,5.1615,0.0052,0.3090,0.0022,0.0002
2,TAIWAN,186,239,957,4.5035,0.0106,0.2304,0.0068,0.0000
1,THAILAND,555,336,2431,5.4629,0.0061,0.2392,0.0034,0.0012
0,UK,386,220,1823,6.0165,0.0099,0.2444,0.0038,0.0000
6,USA,807,600,5157,7.3830,0.0053,0.2867,0.0017,0.0044
\end{filecontents*}
\csvreader[%
before reading=\footnotesize\setlength{\tabcolsep}{1.5pt},
 tabular={|p{2.2cm}|p{1.4cm}|p{1.4cm}|p{1.4cm}|p{1.2cm}|p{1.7cm}|p{1.2cm}|p{1.4cm}|p{1.9cm}|},
        table head = \hline\textbf{Country Name} & \textbf{{Num-Company}} & \textbf{{Num-Customer}} & \textbf{Num-Edges} & \textbf{Average Degree} & \textbf{Clustering Coefficient} & \textbf{Density} & \textbf{Closeness} & \textbf{Betweenness}\\\hline,
        late after line= \\,
        late after last line=\\\hline %
        ]{tables/customer_network_params_new.csv}{country=\country,numHead=\numHead,numTail=\numTail,numEdge=\numEdge,averageDegree=\averageDegree,averageClusteringCoeffi=\averageClusteringCoeffi,density=\density, closeness=\closeness, betweenness=\betweenness}%
        {\country & \numHead & \numTail & \numEdge & \averageDegree & \averageClusteringCoeffi & \density & \closeness & \betweenness}
       \centering
        \caption{Statistic analysis of the \textit{supplies\_to} network that represents the supplying relationship of the company and the customer.}
        \label{tab: supplier to network}
\end{table*}

 \vspace{-5.0pt}

\begin{table*}[ht]
\centering
\begin{filecontents*}{tables/product_network_params_new.csv}
,country,numHead,numTail,numEdge,averageDegree,density,closeness,betweenness,averageClusteringCoeffi
9,BRAZIL,200,341,5163,19.0869,0.0353,0.3884,0.0030,0.0000
5,CHINA,963,868,79209,86.5199,0.0473,0.4033,0.0008,0.0000
4,GERMANY,613,728,51445,76.7263,0.0573,0.4131,0.0011,0.0000
3,INDIA,510,622,27272,48.1837,0.0426,0.4017,0.0013,0.0000
8,JAPAN,980,883,71582,76.8459,0.0413,0.4036,0.0008,0.0000
7,KOREA,291,685,12409,25.4283,0.0261,0.4107,0.0015,0.0000
2,TAIWAN,239,374,6189,20.1925,0.0330,0.3765,0.0028,0.0000
1,THAILAND,336,479,9001,22.0883,0.0271,0.3729,0.0021,0.0000
0,UK,220,433,7630,23.3691,0.0358,0.3835,0.0025,0.0000
6,USA,600,742,41876,62.4083,0.0465,0.4042,0.0011,0.0000
\end{filecontents*}
\csvreader[%
before reading=\footnotesize\setlength{\tabcolsep}{1.5pt},
 tabular={|p{2.2cm}|p{1.4cm}|p{1.4cm}|p{1.4cm}|p{1.2cm}|p{1.7cm}|p{1.2cm}|p{1.4cm}|p{1.9cm}|},
        table head = \hline\textbf{Country Name} & \textbf{{Num-Customer}} & \textbf{{Num-Product}} & \textbf{Num-Edges} & \textbf{Average Degree} & \textbf{Clustering Coefficient} & \textbf{Density} & \textbf{Closeness} & \textbf{Betweenness}\\\hline,
        late after line= \\,
        late after last line=\\\hline %
        ]{tables/product_network_params_new.csv}{country=\country,numHead=\numHead,numTail=\numTail,numEdge=\numEdge,averageDegree=\averageDegree,averageClusteringCoeffi=\averageClusteringCoeffi,density=\density, closeness=\closeness, betweenness=\betweenness}%
        {\country & \numHead & \numTail & \numEdge & \averageDegree & \averageClusteringCoeffi & \density & \closeness & \betweenness}
       \centering
        \caption{Statistic analysis of the \textit{buys} network that represents the buying relationship of the customer and the product.}
        \label{tab: buys network}
\end{table*}

\begin{table*}[!ht]
\centering
\begin{filecontents*}{tables/company_network_params_new.csv}
,country,numHead,numTail,numEdge,averageDegree,density,closeness,betweenness,averageClusteringCoeffi
9,BRAZIL,568,838,4742,6.7454,0.0048,0.2739,0.0019,0.0000
5,CHINA,885,11842,65422,10.2808,0.0008,0.2994,0.0002,0.0000
4,GERMANY,801,2165,10537,7.1052,0.0024,0.2737,0.0009,0.0000
3,INDIA,696,1318,8625,8.5650,0.0043,0.2825,0.0013,0.0000
8,JAPAN,898,6553,30559,8.2027,0.0011,0.2867,0.0003,0.0000
7,KOREA,729,1070,6816,7.5775,0.0042,0.2726,0.0015,0.0000
2,TAIWAN,527,764,4490,6.9558,0.0054,0.2630,0.0021,0.0000
1,THAILAND,599,1963,7821,6.1054,0.0024,0.2616,0.0011,0.0000
0,UK,593,1036,4114,5.0510,0.0031,0.2410,0.0019,0.0000
6,USA,853,3297,16873,8.1316,0.0020,0.2798,0.0006,0.0000
\end{filecontents*}
\csvreader[%
before reading=\footnotesize\setlength{\tabcolsep}{1.5pt},
 tabular={|p{2.2cm}|p{1.4cm}|p{1.4cm}|p{1.4cm}|p{1.2cm}|p{1.7cm}|p{1.2cm}|p{1.4cm}|p{1.9cm}|},
        table head = \hline\textbf{Country Name} & \textbf{{Num-Product}} & \textbf{{Num-Company}} & \textbf{Num-Edges} & \textbf{Average Degree} & \textbf{Clustering Coefficient} & \textbf{Density} & \textbf{Closeness} & \textbf{Betweenness}\\\hline,
        late after line= \\,
        late after last line=\\\hline %
        ]{tables/company_network_params_new.csv}{country=\country,numHead=\numHead,numTail=\numTail,numEdge=\numEdge,averageDegree=\averageDegree,averageClusteringCoeffi=\averageClusteringCoeffi,density=\density, closeness=\closeness, betweenness=\betweenness}%
        {\country & \numHead & \numTail & \numEdge & \averageDegree & \averageClusteringCoeffi & \density & \closeness & \betweenness}
       \centering
        \caption{Statistic analysis of the \textit{made\_by} network that represents the relationship of the product and the company.}
        \label{tab: made by network}
\end{table*}

\begin{table*}[!ht]
\centering
\begin{filecontents*}{tables/certification_network_params_new.csv}
,country,numHead,numTail,numEdge,averageDegree,density,closeness,betweenness,averageClusteringCoeffi
9,BRAZIL,323,5,504,3.0732,0.0094,0.4059,0.0047,0.0000
5,CHINA,7730,5,12704,3.2848,0.0004,0.3882,0.0002,0.0000
4,GERMANY,1150,5,1877,3.2502,0.0028,0.3589,0.0016,0.0000
3,INDIA,821,5,1427,3.4552,0.0042,0.3765,0.0021,0.0000
8,JAPAN,2970,5,3671,2.4679,0.0008,0.4675,0.0004,0.0000
7,KOREA,769,5,1715,4.4315,0.0057,0.4303,0.0018,0.0000
2,TAIWAN,521,5,1029,3.9125,0.0075,0.3987,0.0030,0.0000
1,THAILAND,707,4,1026,2.8861,0.0041,0.3590,0.0026,0.0000
0,UK,622,5,1022,3.2600,0.0052,0.3793,0.0027,0.0000
6,USA,1564,5,2468,3.1460,0.0020,0.3655,0.0011,0.0000
\end{filecontents*}
\csvreader[%
before reading=\footnotesize\setlength{\tabcolsep}{1.5pt},
 tabular={|p{2.2cm}|p{1.4cm}|p{1.4cm}|p{1.4cm}|p{1.2cm}|p{1.7cm}|p{1.2cm}|p{1.4cm}|p{1.9cm}|},
        table head = \hline\textbf{Country Name} & \textbf{{Num-Company}} & \textbf{{Num-Certification}} & \textbf{Num-Edges} & \textbf{Average Degree} & \textbf{Clustering Coefficient} & \textbf{Density} & \textbf{Closeness} & \textbf{Betweenness}\\\hline,
        late after line= \\,
        late after last line=\\\hline %
        ]{tables/certification_network_params_new.csv}{country=\country,numHead=\numHead,numTail=\numTail,numEdge=\numEdge,averageDegree=\averageDegree,averageClusteringCoeffi=\averageClusteringCoeffi,density=\density, closeness=\closeness, betweenness=\betweenness}%
        {\country & \numHead & \numTail & \numEdge & \averageDegree & \averageClusteringCoeffi & \density & \closeness & \betweenness}
       \centering
        \caption{Statistic analysis of the \textit{has\_cert} network that represents the relationship of the company and the certification.}
        \label{tab: has cert network}
\end{table*}






\end{document}